\documentclass[10pt,DIV=12]{scrartcl}
\pdfoutput=1

\usepackage[T1]{fontenc}
\usepackage[utf8]{inputenc}
\newcommand{\unicodebeta}{^^ce^^b2}
\newcommand{\unicodesubscripth}{^^e2^^82^^95}
\newcommand{\unicodemodifierd}{^^e1^^b5^^88}
\newcommand{\unicodemodifierm}{^^e1^^b5^^90}
\newcommand{\unicodemodifiero}{^^e1^^b5^^92}

\usepackage{amsmath,amsfonts}
\usepackage[mathscr]{euscript}
\usepackage{booktabs}
\usepackage{cite}
\usepackage{graphicx, ulem, hhline}
\usepackage[table]{xcolor}
\usepackage{siunitx}
\usepackage{soul}
\usepackage{wasysym}              
\usepackage{hyphenat}             
\usepackage[titles]{tocloft}

\usepackage{etoolbox,xstring,xspace}

\let\theparentequation\theequation
\patchcmd{\theparentequation}{equation}{parentequation}{}{}

\renewenvironment{subequations}[1][]{
  \refstepcounter{equation}%
  \setcounter{parentequation}{\value{equation}}
  \setcounter{equation}{0}
  \def\theequation{\theparentequation\alph{equation}}%
  \let\parentlabel\label
  \ifx\\#1\\\relax\else\label{#1}\fi
  \ignorespaces
}{%
  \setcounter{equation}{\value{parentequation}}
  \ignorespacesafterend
}

\newcommand*{\nextParentEquation}[1][]{
  \refstepcounter{parentequation}
  \setcounter{equation}{0}
  \ifx\\#1\\\relax\else\parentlabel{#1}\fi
}

\makeatletter
\renewcommand\@makefntext[1]{\leftskip=.7em\hskip-.69em\@makefnmark#1}
\makeatother
\usepackage{setspace}

\usepackage{jheppub}
\usepackage{hypernat}

\newrobustcmd*{\tocref}[1]{\hyperref[TOC]{\color{black}{#1}}}
\newcommand{\tocsection}[2][]{\section[\boldmath #2]{\tocref{\boldmath #2#1}}}
\newcommand{\tocsubsection}[2][]{\subsection[#2]{\tocref{\boldmath #2#1}}}

\renewcommand*{\backref}[1]{}
\renewcommand*{\backrefalt}[4]{%
  \ifcase #1%
  \or [p\,#2]%
  \else [pp\,#2]%
  \fi%
}

\newif\ifbackrefshowonlyfirst
\backrefshowonlyfirsttrue
\makeatletter
\let\BR@direct@old@hyper@natlinkstart\hyper@natlinkstart
\renewcommand*{\hyper@natlinkstart}{\phantomsection\BR@direct@old@hyper@natlinkstart}
\let\BR@direct@oldBR@citex\BR@citex
\renewcommand*{\BR@citex}{\phantomsection\BR@direct@oldBR@citex}%

\long\def\hyper@page@BR@direct@ref#1#2#3{\hyperlink{#3}{#1}}

\ifx\backrefxxx\hyper@page@backref
    \let\backrefxxx\hyper@page@BR@direct@ref
    \ifbackrefshowonlyfirst
    \fi
\else
    \ifbackrefshowonlyfirst
    \fi
\fi

\patchcmd{\Hy@backout}{Doc-Start}{\@currentHref}{}{\errmessage{I can't seem to patch backref}}
\makeatother

\usepackage[all]{hypcap}
\usepackage{footnotebackref}

\usepackage[format=plain, margin=0.cm, font=small, labelfont=bf]{caption}
\DeclareCaptionSubType*[arabic]{figure}
\DeclareCaptionSubType*[arabic]{table}
\let\theparentequation\theequation
\patchcmd{\theparentequation}{equation}{parentequation}{}{}

\apptocmd{\thebibliography}{\small}{}{}
\let\OLDthebibliography\thebibliography
\renewcommand\thebibliography[1]{
  \OLDthebibliography{#1}
  \setlength{\parskip}{1pt}
  \setlength{\itemsep}{1pt plus 0.3ex}
}

\def\unity{\textbf{1}}

\newcommand{\IE}{\textit{i.\,e.}}
\newcommand{\EG}{\textit{e.\,g.}}
\newcommand{\I}{{i\mkern1mu}}
\newcommand{\E}{\mathrm{e}}
\newcommand{\Real}[1]{\Re\hspace{-1pt}\mathfrak{e}{\left[#1\right]}}

\newcommand{\MZ}{\mathbf{M}^{\mathbf{Z}}}

\newcommand{\Bnull}[1]{B_{0}{\left(#1\right)}}

\newcommand{\Beins}[1]{B_{1}{\left(#1\right)}}

\newcommand{\mathsym}[1]{{}}
\newcommand{\unicode}[1]{{}}

\BeforeTOCHead[toc]{{\pdfbookmark[1]{\contentsname}{toc}}}

\newcommand*{\mytitle}[1]{%
  \parbox{\linewidth}{\setstretch{1.5}\centering\Large\textsc{\textbf{\boldmath #1}}}
}

\hypersetup{pdfauthor={Sophia Borowka, Sebastian Pa{\ss}ehr and Georg Weiglein},
            pdftitle={Complete two-loop QCD contributions to the lightest Higgs-boson mass in the MSSM with complex parameters}}

\begin{document}


\thispagestyle{empty}

\def\thefootnote{\fnsymbol{footnote}}

\begin{flushright}
DESY--17--066 \\
CERN--TH--2017--081
\end{flushright}

\vspace{2cm}

\begin{center}

\mytitle{
Complete two-loop QCD contributions to the\\
lightest Higgs-boson mass in the MSSM\\
with complex parameters
}

\vspace{1cm}

Sophia Borowka$^1$\footnote{email: sophia.borowka@cern.ch}, Sebastian Pa{\ss}ehr$^{2,3}$\footnote{email: passehr@lpthe.jussieu.fr}
and
Georg Weiglein$^3$\footnote{email: georg.weiglein@desy.de}

\vspace*{.7cm}

\textsl{
$^1$ Theoretical Physics Department, CERN, Geneva, Switzerland
}

\medskip
\textsl{
$^2$Sorbonne Université, CNRS,
Laboratoire de Physique Théorique et Hautes Énergies (LPTHE),\\
4 Place Jussieu, F--75252 Paris CEDEX~05, France
}

\medskip
\textsl{
$^3$Deutsches Elektronensynchrotron DESY,\\
Notkestraße 85, D--22607 Hamburg, Germany
}

\end{center}

\vspace*{2cm}

\begin{abstract}{}
Higher-order corrections to the MSSM Higgs-boson masses are desirable
for accurate predictions currently testable at the LHC.  By comparing
the prediction with the measured value of the discovered Higgs signal,
viable parameter regions can be inferred. For an improved theory
accuracy, we compute all two-loop corrections involving the strong
coupling for the Higgs-boson mass spectrum of the MSSM with complex
parameters. Apart from the dependence on the strong coupling, these
contributions depend on the weak coupling and Yukawa couplings,
leading to terms of~$\mathcal{O}{\left(\alpha\alpha_s\right)}$
and~$\mathcal{O}{\left(\sqrt{\alpha_{q_1}}\sqrt{\alpha_{q_2}}\alpha_s\right)}$,
\mbox{($q_{1,2}=t,b,c,s,u,d$)}. The full dependence on the external
momentum and all relevant mass scales is taken into account. The
calculation is performed in the Feynman-diagrammatic approach which is
flexible in the choice of the employed renormalization scheme. For the
phenomenological results presented here, a renormalization scheme
consistent with higher-order corrections included in the
code~\texttt{FeynHiggs} is adopted. For the evaluation of the results,
a total of 513 two-loop two-point integrals with up to five different
mass scales are computed fully numerically using the
program~\texttt{SecDec}. A comparison with existing results in the
limit of real parameters and/or vanishing external momentum is carried
out, and the impact on the lightest Higgs-boson mass is discussed,
including the dependence on complex phases. The new results will be
included in the public code~\texttt{FeynHiggs}.
\end{abstract}

\def\thefootnote{\arabic{footnote}}
\setcounter{page}{0}
\setcounter{footnote}{0}

\newpage
\hypersetup{linkcolor=black}
\tableofcontents\label{TOC}
\hypersetup{linkcolor=blue}

\tocsection{Introduction}

Since the discovery of a signal in the Higgs-boson searches at the
LHC~\cite{Aad:2012tfa,Chatrchyan:2012ufa} with a mass around
$125$~GeV, it is a prime goal to reveal the detailed nature of the new
particle.  While with the present experimental and theoretical
uncertainties the measured properties of the detected particle are
compatible with the expectations for the Higgs boson of the Standard
Model (SM)~\cite{Sperka:2018,Khachatryan:2016vau}, other
interpretations corresponding to very different underlying physics are
also in agreement with the data. A crucial question in this context is
in particular whether the observed particle is part of an extended
Higgs sector that would be associated with a more general theoretical
framework beyond the SM.

\medskip 

Within the theoretically well motivated Minimal Supersymmetric
extension of the SM~(MSSM), the observed particle can be interpreted
as a light state within a richer spectrum of scalar particles.%
\footnote{Within the MSSM it is usually assumed that the observed
  particle is associated with the lightest neutral Higgs boson of the
  model; see Ref.~\cite{Bechtle:2016kui} for a recent update on the
  viability of the interpretation in terms of the next-to-lightest
  neutral Higgs boson of the MSSM.}  The Higgs-boson sector of the
MSSM consists of two complex scalar doublets leading to five physical
Higgs bosons and three (would-be) Goldstone bosons.  At the
tree-level, the physical states are given by the  neutral
$CP$-even bosons $h,H$ and the $CP$-odd state $A$, together with the
charged $H^{\pm}$ bosons. The Higgs sector at lowest order  can be
parametrized in terms of the $A$-boson mass $m_A$ and the ratio of the
vacuum expectation values of  the scalar doublets, $\tan\beta =
\left.v_2\middle/v_1\right.$. The MSSM with complex parameters~(cMSSM)
is of particular interest since it provides new sources of
$CP$-violation in addition to the $CP$-violating phase of the SM.
Thereby the Higgs sector is $CP$-conserving at the tree level, but
potentially large loop contributions involving complex parameters from
other supersymmetric (SUSY) sectors can lead to an admixture of the
$CP$-even states $h,\,H,$ and the $CP$-odd $A$ resulting in the mass
eigenstates $h_1, h_2,
h_3$~\cite{Pilaftsis:1998pe,Demir:1999hj,Pilaftsis:1999qt,
Heinemeyer:2001qd,Frank:2006yh}. In this case $m_A$ is no longer a
useful input parameter; instead the mass of the charged Higgs boson
$m_{H^\pm}$ is used.  Besides the input parameter $m_A$ or $m_{H^\pm}$
all other Higgs-boson masses are predicted quantities in the MSSM. The
Higgs-boson masses and mixings in the neutral sector are strongly
affected by loop contributions. Especially for the experimentally
measured Higgs boson at about $125$~GeV a sufficiently high accuracy
of the theoretical computation is essential for drawing reliable
conclusions on the viability of the investigated region of parameter
space.

A large amount of work has been invested into calculating higher-order
corrections to the mass spectrum within the MSSM with real
parameters~\cite{Degrassi:2002fi,Heinemeyer:1998np,Heinemeyer:1998jw,Heinemeyer:1999be,Haber:1990aw,Ellis:1990nz,Okada:1990vk,Okada:1990gg,Ellis:1991zd,Sasaki:1991qu,Chankowski:1991md,Brignole:1992uf,Hempfling:1993qq,Casas:1994us,Dabelstein:1994hb,Carena:1995bx,Carena:1995wu,Pierce:1996zz,Haber:1996fp,Heinemeyer:1998kz,Zhang:1998bm,Espinosa:1999zm,Carena:2000dp,Espinosa:2000df,Espinosa:2001mm,Degrassi:2001yf,Martin:2001vx,Brignole:2001jy,Dedes:2002dy,Martin:2002iu,Dedes:2003km,Heinemeyer:2004xw,Martin:2003qz,Martin:2003it,Allanach:2004rh,Heinemeyer:2004gx,Martin:2005eg,Martin:2005qm,Harlander:2008ju,Kant:2010tf,Brignole:2002bz,Harlander:2017kuc,Borowka:2014wla,Degrassi:2014pfa,Borowka:2015ura,Draper:2013oza,Vega:2015fna,Lee:2015uza,Hahn:2013ria,Bahl:2016brp,Bahl:2017aev,Athron:2016fuq,Bagnaschi:2014rsa,Bagnaschi:2017xid,Martin:2002wn,Martin:2004kr,Martin:2007pg}
as well as the MSSM with complex
parameters~\cite{Pilaftsis:1998pe,Demir:1999hj,Pilaftsis:1999qt,Choi:2000wz,Ibrahim:2000qj,Ibrahim:2002zk,Carena:2000yi,Heinemeyer:2001qd,Martin:2002wn,Martin:2004kr,Frank:2006yh,Martin:2007pg,Heinemeyer:2007aq,Hollik:2014wea,Hollik:2014bua,Passehr:2017ufr,Goodsell:2016udb}.
The largest loop contributions originate from the Yukawa sector due to
the size of the top-quark Yukawa coupling~$h_t$,
where \mbox{$\alpha_t=\left.h_t^2\middle/(4\pi)\right.$}. At the
two-loop level QCD corrections enter. The dominant contribution at the
two-loop level is given by
the~$\mathcal{O}{\left(\alpha_{t}\alpha_{s}\right)}$ terms which are
known for the MSSM with complex
parameters~\cite{Heinemeyer:1998jw,Heinemeyer:1998np,Heinemeyer:1999be,
Heinemeyer:2007aq}. Also
the~$\mathcal{O}{\left(\alpha_{t}^2+\alpha_t\alpha_b+\alpha_b^2\right)}$
corrections are known for the case of complex
parameters~\cite{Hollik:2014wea,Hollik:2014bua,Passehr:2017ufr}.
Restricting to the case of real parameters, the momentum-dependent
$\mathcal{O}{\left(\alpha_{t}\alpha_{s}\right)}$
corrections~\cite{Borowka:2014wla,Degrassi:2014pfa,Borowka:2015ura}
and the contributions for the case where all Yukawa couplings except
the one of the top quark are neglected~\cite{Degrassi:2014pfa} are
known.  While the phases of the complex parameters affect the
predictions for the Higgs-boson masses, production cross
sections~\cite{Liebler:2016ceh} and
decays~\cite{Heinemeyer:1998yj,Lee:2003nta,Williams:2011bu}, they also
induce $CP$-violating effects that are constrained by other
experiments. These concern in particular the electric dipole
moments~\cite{Baron:2013eja,Pospelov:2005pr,Afach:2015sja,Baker:2006ts,Baker:2007df,Serebrov:2015idv,Graner:2016ses,Yamanaka:2017mef}.
For the usual convention where the phase of the mass of the
electroweakinos, $\phi_{M_2}$, is set to zero without loss of
generality, the phase of the parameter $\mu$ is constrained to be very
close to zero or $\pi$.  The other important phases of the gluino
mass, $\phi_{M_3}$, and the trilinear soft-breaking parameters of the
stops, $\phi_{A_t}$, and sbottoms, $\phi_{A_b}$, are much less
constrained.  In particular, the bounds on the phases of the trilinear
soft-breaking parameters are significantly weaker for the third
generation than for the second and first generation.

\medskip 

In this article the full two-loop QCD corrections to the Higgs-boson
masses are presented for the general case of the MSSM with complex
parameters. They contain all previously computed results for the MSSM
with real or complex parameters. The contributions are comprised of
the $\mathcal{O}{\left(\alpha\alpha_s\right)}$ terms, involving the
electroweak gauge coupling $\alpha$, and the
$\mathcal{O}{\left(\sqrt{\alpha_{q_1}}\sqrt{\alpha_{q_2}}\alpha_s\right)}$
terms, involving the Yukawa couplings $\alpha_{q_1}$, $\alpha_{q_2}$,
where $q_{1,2}=t,b,c,s,u,d$. Terms with mixed up- and down-type Yukawa
couplings only appear in conjunction with $m_{H^\pm}$ as input
parameter. Mixed contributions of
$\mathcal{O}{\left(\sqrt{\alpha}\sqrt{\alpha_{q_1}}\alpha_s\right)}$
involving one gauge coupling and one Yukawa coupling do not appear in
the final result.  The results obtained here for the MSSM can
furthermore be used as an approximation for higher-order contributions
within the~NMSSM, as discussed in
Refs.~\cite{Drechsel:2016jdg,Drechsel:2016htw,Domingo:2017rhb}.  The
computation carried out below makes use of previously developed
tools~\cite{Weiglein:1993hd,Borowka:2014wla,Hahn:2015gaa}.  The
momentum-dependent two-loop integrals appearing in the two-loop QCD
corrections are evaluated with an adapted version of
\texttt{SecDec 2}~\cite{Carter:2010hi,Borowka:2012yc,Borowka:2013cma}.  
For the numerical analysis the new contributions are combined with the
full one-loop result~\cite{Frank:2006yh} and the
leading~$\mathcal{O}{\left(\alpha_{t}^2\right)}$
terms~\cite{Hollik:2014wea,Hollik:2014bua} in the Feynman-diagrammatic
approach for complex parameters, available through the public
program \texttt{FeynHiggs}~\cite{Heinemeyer:1998np,Degrassi:2002fi,Frank:2006yh,Heinemeyer:1998yj,Hahn:2010te}.
In deriving the new contributions the renormalization scheme of
Ref.~\cite{Frank:2006yh} at the one-loop level has been adopted and
applied to the case of the~$\mathcal{O}{\left(\alpha\alpha_s\right)}$
contributions.  This ensures that the obtained analytical results for
the renormalized two-loop self-energies can consistently be
incorporated into \texttt{FeynHiggs}.  In the results presented in
this paper no resummation of higher-order logarithmic contributions as
obtained in
Refs.~\cite{Hahn:2013ria,Draper:2013oza,Vega:2015fna,Lee:2015uza,Bahl:2016brp,Athron:2016fuq,Bahl:2017aev}
has been included. The combination of resummed higher-order
logarithmic contributions with the results obtained in the present
paper will be addressed in future work.  In the numerical analysis
below, we show results for the masses of the three neutral Higgs
bosons of the MSSM with complex parameters and their phase dependence,
with a particular focus on those results which are phenomenologically
most relevant.

\medskip 

The paper is organized as follows: section~\ref{sec:HiggsSect}
provides the theoretical framework for the calculation and
renormalization of the Feynman diagrams that is used to arrive at
expressions for the dressed propagators of the Higgs sector up to the
two-loop level.  The calculation of the unrenormalized self-energies
and the construction of the two-loop counterterms are described in
section~\ref{sec:thecalculation}.  Details on the numerical evaluation
of the momentum-dependent two-loop integrals are given in
section~\ref{sec:numericeval}, whereas the impact of the new
contributions on the Higgs-boson masses is discussed in
section~\ref{sec:numeric}.  The conclusions are given in
section~\ref{sec:concl}.

\tocsection[\label{sec:HiggsSect}]{The Higgs sector of the MSSM with complex parameters}

\tocsubsection{Tree-level relations for masses and mixing}

The two scalar $SU(2)$-doublets are conventionally expressed in terms
of their components as follows,
\begin{alignat}{2}
  \label{eq:Higgsfields}
  \mathcal{H}_{1} &= \begin{pmatrix} v_{1} + \frac{1}{\sqrt{2}}(\phi_{1} - \I \chi_{1})\\ -\phi^{-}_{1}\end{pmatrix},&\quad
  \mathcal{H}_{2} &= \E^{\I \xi}\begin{pmatrix} \phi^{+}_{2}\\ v_{2} + \frac{1}{\sqrt{2}}(\phi_{2} + \I \chi_{2})\end{pmatrix} \,,
\end{alignat}
with the relative phase $\xi$. The Higgs potential can be written as a
polynomial in the field components,
\begin{align}
  \begin{split}
    V_{H} &= -T_{\phi_{1}} \phi_{1} - T_{\phi_{2}} \phi_{2} - T_{\chi_{1}} \chi_{1} - T_{\chi_{2}} \chi_{2}\\
         &\quad + \frac{1}{2}\begin{pmatrix} \phi_{1}, & \phi_{2}, & \chi_{1}, & \chi_{2} \end{pmatrix}
            \begin{pmatrix}\mathbf{M}_{\phi} & \mathbf{M}_{\phi\chi}\\ \mathbf{M}_{\phi\chi}^{\dagger} & \mathbf{M}_{\chi} \end{pmatrix}
         \begin{pmatrix} \phi_{1}, & \phi_{2}, & \chi_{1}, & \chi_{2}\end{pmatrix}^{\text{T}}
            + \begin{pmatrix} \phi^{-}_{1}, & \phi^{-}_{2}\end{pmatrix} \mathbf{M}_{\phi^{\pm}} \begin{pmatrix} \phi^{+}_{1}\\ \phi^{+}_{2}\end{pmatrix} + \dots  \ ,
\label{eq:VH}
  \end{split}
\end{align}
where terms of third and fourth power in the fields have been omitted,
and the relations~$\phi^{-}_{1} =
\left(\phi^{+}_{1}\right)^{\dagger}$ and~$\phi^{-}_{2} =
\left(\phi^{+}_{2}\right)^{\dagger}$ have been used.  Explicit
expressions for the tadpole coefficients $T$ and for the mass matrices
$\mathbf{M}$ can be found in Ref.~\cite{Frank:2006yh}. They are
parametrized by the phase~$\xi$, the real SUSY-breaking quantities
\mbox{$m_{1,2}^{2} = \tilde{m}_{1,2}^{2} + \lvert\mu\rvert^{2}$}, and
the complex SUSY-breaking quantity~$m_{12}^{2}$. With the help of a
Peccei--Quinn transformation~\cite{Peccei:1977hh} the
parameter~$m_{12}^{2}$ can be redefined such that its phase
vanishes~\cite{Dimopoulos:1995kn}, leaving only the phase~$\xi$ as a
potential source of $CP$-violation at tree level. The requirement of
minimizing~$V_{H}$ at the vacuum expectation values~$v_{1}$
and~$v_{2}$ is equivalent to the requirement of vanishing tadpoles of
the physical fields, which in turn implies the condition~$\xi = 0$ at
tree level.  As a consequence, the Higgs sector of the MSSM is
$CP$-conserving at lowest order. This implies in Eq.~\eqref{eq:VH}
that~$\mathbf{M}_{\phi\chi}$ is equal to zero, and~$\phi_{1,2}$ do not
mix with~$\chi_{1,2}$ at tree-level.

The remaining $(2\times 2)$-matrices $\mathbf{M}_{\phi}$,
$\mathbf{M}_{\chi}$, $\mathbf{M}_{\phi^{\pm}}$ can be transformed into
the mass eigenstate basis with the help of orthogonal matrices~$D(x)$,
using the
abbreviations \mbox{$s_{x} \equiv \sin{x}$}, \mbox{$c_{x} \equiv \cos{x}$},
\begin{alignat}{8}
  \label{eq:higgsmixing}
  D{\left(x\right)} &= \begin{pmatrix}-s_{x} & c_{x}\\ c_{x} & s_{x}\end{pmatrix},&\quad
  \begin{pmatrix} h\\ H \end{pmatrix} &= D(\alpha) \begin{pmatrix} \phi_{1}\\ \phi_{2}\end{pmatrix},&\quad
  \begin{pmatrix} A\\ G \end{pmatrix} &= D(\beta_{n}) \begin{pmatrix} \chi_{1}\\ \chi_{2}\end{pmatrix},&\quad
  \begin{pmatrix} H^{\pm}\\ G^{\pm} \end{pmatrix} &= D(\beta_{c}) \begin{pmatrix} \phi^{\pm}_{1}\\ \phi^{\pm}_{2}\end{pmatrix}.
\end{alignat}
The Higgs potential in this basis can be expressed as follows,
\begin{align}
\label{eq:HiggsPotential}
  \begin{split}
    V_{H} &= -T_h \, h- T_H \, H - T_A \, A - T_G\,  G\\
         &\quad + \frac{1}{2}\begin{pmatrix} h, & H, & A, & G \end{pmatrix}
    \mathbf{M}_{hHAG}
    \begin{pmatrix} h, & H, & A, & G \end{pmatrix}^{\text{T}}
            + \begin{pmatrix} H^{-}, & G^{-}\end{pmatrix} 
            \mathbf{M}_{H^\pm G^\pm}
                \begin{pmatrix} H^{+}\\ G^{+}\end{pmatrix}
            + \dots\ 
  \end{split}
\end{align}
with the tadpole coefficients $T_{h,H,A,G}$ and the mass matrices
\begin{align}
\label{eq:mmatrices}
\mathbf{M}_{hHAG} &= \begin{pmatrix}
        m^2_{h} & m^2_{hH} & m^2_{hA} & m^2_{hG} \\
        m^2_{hH} & m^2_{H} & m^2_{HA} & m^2_{HG} \\
        m^2_{hA} & m^2_{HA} & m^2_{A} & m^2_{AG} \\
        m^2_{hG} & m^2_{HG} & m^2_{AG} & m^2_{G} \end{pmatrix} , \qquad
\mathbf{M}_{H^\pm G\pm} \,= \begin{pmatrix}
        m^2_{H^\pm}  &  m^2_{H^-G^+} \\
        m^2_{G^-H^+} &  m^2_{G^\pm} \end{pmatrix}  ;
\end{align} 
explicit expressions for the entries are given in
Ref.~\cite{Frank:2006yh}. The tadpole terms in
Eq.~(\ref{eq:HiggsPotential}) are zero at the tree level, but they
enter the predictions for the Higgs-boson masses at higher orders. As
mentioned above, the ellipses denote terms of higher power in the
fields which are not relevant in our calculation.

After applying the minimization conditions to
Eqs.~(\ref{eq:mmatrices}), the mass matrices can be brought into
canonical form\footnote{We use a lower-case~$m$ for the Higgs-boson
masses at the tree level.}
\begin{align}
\label{eq:mmatricesdiag}
\mathbf{M}_{ h H A G}^{(0)}  & =\, \mathrm{diag} \left( m_h^2,\, m_H^2,\, m_A^2,\, m_G^2 \right) , \quad
\mathbf{M}_{H^\pm G^\pm}^{(0)} \, =\,  \mathrm{diag} \left( m_{H^\pm}^2,\, m_{G^\pm}^2 \right) ,
\end{align}
for $\beta = \beta_{n} = \beta_{c}$, with $\beta \in
\left[0,\pi/2\right)$ given in terms of the vacuum expectation values,
\begin{align}
\label{eq:deftanbeta}
\tan\beta &\equiv t_\beta = \frac{v_2}{v_1} \ , 
\end{align}
and for the second mixing angle $\alpha \in \left[-\pi/2,0\right)$
determined by
\begin{align}
  \label{eq:alpha}
  \tan (2\alpha) &= \frac{m_{A}^2 + m_Z^2}{m_A^2 - m_Z^2}\, \tan(2\beta)\, .
\end{align}
The Goldstone bosons are massless\footnote{The Goldstone bosons can
  acquire a non-zero mass value by gauge fixing.}, $m_{G^\pm} = m_G =
  0$. The masses $m_{H^\pm}, m_A, m_h, m_H$ fullfil the relations%
\begin{subequations}
\label{eq:treelevelmasses}
\begin{align}
  \label{subeq:mhpmma}
  m_{H^{\pm}}^{2} &= m_{A}^{2} + M_{W}^{2} \, , \\
  m_{h,\,H}^2 &= \frac{1}{2}\left(m_{A}^{2} +
    M_{Z}^{2}\mp\sqrt{\left(m_{A}^{2} + M_{Z}^{2}\right)^{2} - 4 m_{A}^{2} M_{Z}^{2}\, c_{2\beta}^{2}}\right) ,
\end{align}
\end{subequations}
including the vector-boson masses $M_W$ and $M_Z$. Given the relation
in Eq.~(\ref{subeq:mhpmma}), both $m_A$ and $m_{H^{\pm}}$ can be
chosen as input parameter.

At lowest order, the irreducible two-point vertex functions of the
neutral Higgs sector
\begin{align}
\label{eq:irredgammazero}
\Gamma^{(0)}_{hHAG}(p^2) & = \, i \, \Big[ p^2 \unity - \mathbf{M}^{(0)}_{hHAG} \Big] 
\end{align}
are diagonal, and the entries of the mass matrices in
Eq.~\eqref{eq:mmatricesdiag} provide the poles of the diagonal
lowest-order propagators
\begin{align}
\Delta^{(0)}_{hHAG} (p^2) &=\, - \Big[ \Gamma^{(0)}_{hHAG}(p^2) \Big]^{-1} \, .
\end{align}

\tocsubsection{Masses and mixing beyond lowest order}

Going beyond leading order, the irreducible two-point functions are dressed
by adding the matrix~$\mathbf{\hat{\Sigma}}_{h H A G}$ of the
renormalized diagonal and non-diagonal self-energies for the $h,H, A,
G$ fields up to the considered order,
\begin{align}
\label{eq:homassmatrix}
 p^2 \unity - \mathbf{M}^{(0)}_{hHAG} & \quad \to \quad
 p^2 \unity - \mathbf{M}^{(0)}_{hHAG}  +\mathbf{\hat{\Sigma}}_{hHAG}(p^2) \; \equiv \;
p^2 \unity - \mathbf{M}_{hHAG}(p^2) \, ,
\end{align}
yielding the full renormalized two-point vertex function
\begin{align}
  \hat{\Gamma}_{hHAG}(p^2) =  \, i \, \Big[ p^2 \unity - \mathbf{M}_{hHAG} \Big]\;.
\end{align}
The latter generally contains a mixing of all fields with equal
quantum numbers.  The dressed propagators are obtained by inverting
the matrix $\hat{\Gamma}_{hHAG}(p^2)$.

Truncating the perturbative expansion at the two-loop level, the
momentum-dependent corrections to the neutral Higgs-boson mass
matrices in Eq.~\eqref{eq:homassmatrix} are given by
\begin{align}
  \label{eq:masscorr}
    \mathbf{M}_{ h H A G}^{(2)} (p^2) &= \mathbf{M}_{ h H A G}^{(0)} - \mathbf{\hat{\Sigma}}_{h H A G}^{(1)}(p^2) - \mathbf{\hat{\Sigma}}_{h H A G}^{(2)}(p^2) \ .
\end{align}

For the MSSM with complex parameters, the one-loop self-energies are
completely known~\cite{Frank:2006yh}, and the leading two-loop
$\mathcal{O}{\left(\alpha_t \alpha_s\right)}$ and
$\mathcal{O}{\left(\alpha_{t}^2+\alpha_t\alpha_b+\alpha_b^2\right)}$
contributions have been obtained in the approximation of zero external
momentum~\cite{Heinemeyer:2007aq,Hollik:2014wea,Hollik:2014bua,Passehr:2017ufr}.
In the case of the MSSM with real parameters also the
momentum-dependent corrections
of~$\mathcal{O}{\left(\alpha_{t}\alpha_{s}\right)}$ are
known~\cite{Borowka:2014wla,Degrassi:2014pfa,Borowka:2015ura}. The
remaining QCD contributions at the two-loop level are completed within
this paper. These contributions comprise terms of the
$\mathcal{O}{\left(\alpha_x \alpha_s\right)}$, where $\alpha_x$ is
either the gauge coupling $\alpha$ or the Yukawa coupling $\alpha_q$
with $q = \{u,d,s,c,b,t\}$. We neglect CKM mixing for those
contributions.

In order to obtain the physical Higgs-boson masses from the dressed
propagators at the considered order, it is sufficient to explicitly
derive the entries of the $(3\times 3)$-submatrix of
Eq.~\eqref{eq:masscorr} corresponding to the $(h,H,A)$-components. A
mixing of the neutral Higgs bosons with the Goldstone boson, as well
as Goldstone--$Z$ mixing, yields subleading two-loop contributions to
the Higgs-boson masses that are not
of~$\mathcal{O}{\left(\alpha_x \alpha_s\right)}$.

The masses of the three neutral Higgs bosons are obtained from the
real parts of the complex poles of the $(h,H,A)$-propagator
matrix. They are obtained as the zeroes of the determinant of the
renormalized two-loop two-point vertex function,\footnote{We use an
uppercase~$M$ for the Higgs masses at higher order.}
\begin{alignat}{4}
  \label{eq:higgspoles}
   \left.\operatorname{det}\hat{\Gamma}^{(2)}_{hHA}{\left(p^2\right)}\right|_{p^2\,=\,M_j^2\,-\,\I\,M_j\,\Gamma_j} &= 0, &\quad
   \hat{\Gamma}^{(2)}_{hHA}{\left(p^2\right)} &= \I \left[p^2 {\unity} - \mathbf{M}_{hHA}^{(2)}{\left(p^2\right)}\right], &\quad
   j \in \{h,H,A\}\,,
\end{alignat}
with $\mathbf{M}_{hHA}^{(2)}$ being the corresponding $(3\times
3)$-submatrix of Eq.~\eqref{eq:masscorr}. The impact of the
self-energies on the mixing and couplings of the various Higgs bosons
to other (MS)SM particles can be obtained with the same formalism as
described in Refs.~\cite{Frank:2006yh,Fuchs:2016swt}.

\tocsection[\label{sec:thecalculation}]{Calculation of the renormalized two-loop self-energies}

The renormalized two-loop self-energies can be written as
\begin{align}
\label{eq:renselfenergies}
\mathbf{\hat{\Sigma}}_{hHA} ^{(2)} (p^2)  
   & =
\mathbf{\Sigma}_{hHA} ^{(2)} (p^2) - \delta^{(2)} \MZ_{hHA} \, ,
\end{align}
where $\mathbf{\Sigma}_{hHA} ^{(2)}$ denotes the unrenormalized
self-energies corresponding to the sum of genuine two-loop diagrams
and one-loop diagrams with counterterm insertions. The symbol
$\delta^{(2)}\MZ_{hHA}$ comprises all two-loop counterterms resulting
from parameter and field renormalization.

\begin{figure}[t!]
  \centering
  \includegraphics[width=.85\linewidth]{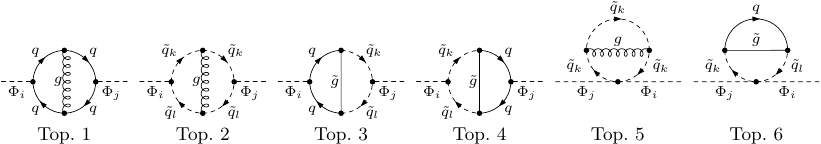}\\[-2ex]
  \rule{.85\linewidth}{\arrayrulewidth}
  \includegraphics[width=.85\linewidth]{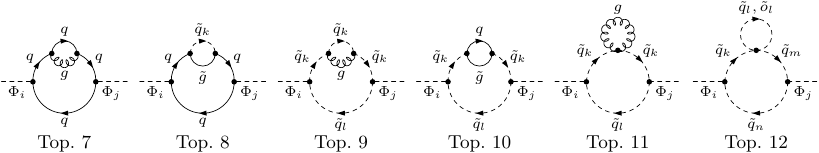}\\[-2ex]
  \rule{.85\linewidth}{\arrayrulewidth}
  \includegraphics[width=.425\linewidth]{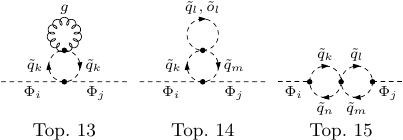}\\[-2ex]
  \rule{.85\linewidth}{\arrayrulewidth}
  \includegraphics[width=.85\linewidth]{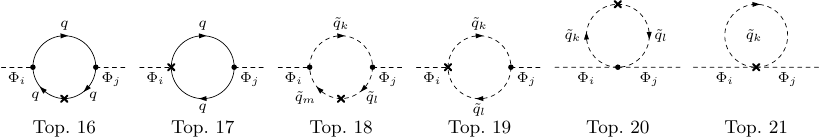}
  \caption{\label{fig:selfenergiesneutral} Types of two-loop
    self-energy diagrams for the neutral Higgs bosons.  One-loop
    counterterm insertions are denoted by a cross.  \mbox{$\Phi_{i} =
      h,\,H,\,A$}; \mbox{$\;\tilde{q} \neq
      \tilde{o}$}. Topologies~$11$ and~$13$ contain a one-point
    loop with a mass-less gluon and are therefore equal to zero.}
  \vspace{-1.1ex}
\end{figure}

The contributing types of Feynman diagrams for the calculation of the
full two-loop QCD corrections entering Eq.~\eqref{eq:renselfenergies}
are depicted in Fig.~\ref{fig:selfenergiesneutral}. The diagrams of
the topologies~$12$, $14$ and~$15$ contribute only if all squarks have
the same flavor; couplings with different flavors vanish since the
color sum is equal to zero in that case. The diagrammatic calculation
has been performed with the help
of~\texttt{FeynArts}~\cite{Kublbeck:1990xc,Hahn:2000kx} in generating
the Feynman diagrams, and \texttt{TwoCalc}~\cite{Weiglein:1993hd}
and~\texttt{Reduze}~\cite{vonManteuffel:2012np} for the two-loop trace
evaluation and tensor reduction. The one-loop renormalization
constants have been computed with the help
of~\texttt{FormCalc}~\cite{Hahn:1998yk}.

\tocsubsection{Two-loop counterterms}

In order to obtain the renormalized self-energies in
Eq.~\eqref{eq:renselfenergies}, counterterms have to be introduced up
to second order in the loop expansion, for the tadpoles
\begin{align}
\label{eq:tadpolct}
  T_i  &\rightarrow  T_i + \delta^{(1)}T_i + \delta^{(2)} T_i \, ,  \quad  i=h,\,H,\,A\,  ,
\end{align}
and for the mass matrices of Eq.~\eqref{eq:HiggsPotential}  
{\allowdisplaybreaks
  \begin{subequations}
    \label{subeqs:counterterms}
\begin{align}  
\mathbf{M}_{ h H A}\label{eq:counterterms}
  &\rightarrow
  \mathbf{M}_{ h H A}^{(0)} + \delta^{(1)} \mathbf{M}_{ h H A} + \delta^{(2)} \mathbf{M}_{ h H A}\ ,  \\
  \label{eq:dm2Lneutral}
  \delta^{(k)}\mathbf{M}_{hHA} &= \begin{pmatrix}\delta^{(k)}m_{h}^{2} & \delta^{(k)}m_{hH}^{2} & \delta^{(k)}m_{hA}^{2} \\
    \delta^{(k)}m_{Hh}^{2} & \delta^{(k)}m_{H}^{2} & \delta^{(k)}m_{HA}^{2}\\
    \delta^{(k)}m_{Ah}^{2} & \delta^{(k)}m_{AH}^{2} & \delta^{(k)}m_{A}^{2}\end{pmatrix},\\
  m^2_{H^\pm}  
  &\rightarrow
  m_{H^\pm}^{2}  +\, \delta^{(1)} m^2_{H^\pm}   
  +\, \delta^{(2)} m^2_{H^\pm}   \, .    
  \label{eq:dm2Lcharged}
\end{align}
\end{subequations}
}%
The two-loop counterterms of
$\mathcal{O}{\left(\alpha \alpha_s\right)}$ have the same structure as
the corresponding one-loop counterterms. They are listed here for
completeness and to fix our notation.

In order to ensure the correct form of the counterterms for the mass
matrices, the rotation angles~$\beta_n$ and $\beta_c$ from
Eqs.~\eqref{eq:higgsmixing} have to be distinguished from $\beta$ in
Eq.~\eqref{eq:deftanbeta}.  Whereas no renormalization is needed for
$\alpha$, $\beta_n$ and $\beta_c$, a counterterm associated with
$\beta$ of the form~$\beta \to \beta +\delta\beta$ is required, in
accordance with the renormalization of $\tan\beta$,
\begin{align}
  \label{eq:tanbetarenormalization}
  t_\beta &\rightarrow t_\beta + \delta^{(1)}t_\beta + \delta^{(2)}t_\beta\ .
\end{align}
In the resulting expressions for the counterterm matrices, the
identification $\beta_c =\beta_n = \beta$ can be made, see
Ref.~\cite{Frank:2006yh} for details of the analogous treatment at the
one-loop order (note that a different convention for the counterterm
of $t_\beta$ is used in Ref.~\cite{Frank:2006yh}).  A complete list of
the two-loop counterterms is given in the Appendix of
Ref.~\cite{Hollik:2014bua}.

In addition to the parameter renormalization described previously, the
field-renormalization constants~$\delta^{(1)}Z_{\mathcal{H}_{i}}$
and~$\delta^{(2)}Z_{\mathcal{H}_{i}}$ are introduced at the one-loop
and two-loop order (restricting the latter to the contributions of
$\mathcal{O}{\left(\alpha \alpha_s\right)}$) for each of the scalar
doublets of Eqs.~\eqref{eq:Higgsfields} through the transformation
\begin{align}\label{eq:twoloopzfactorcontrib}
  \mathcal{H}_{i} \rightarrow \sqrt{Z_{\mathcal{H}_{i}}}\mathcal{H}_{i} &= \left[ 1 + \frac{1}{2}\delta^{(1)}Z_{\mathcal{H}_{i}} + \frac{1}{2}\delta^{(2)}Z_{\mathcal{H}_{i}}
  \right]   \mathcal{H}_{i} \,.
\end{align}
The field-renormalization constants in the mass-eigenstate basis of
Eqs.~\eqref{eq:higgsmixing} are obtained via
\begin{subequations}
\label{eq:neutralhiggsfieldren}
\begin{alignat}{4}
  \begin{pmatrix}h \\ H \end{pmatrix} &\rightarrow
  D(\alpha)\begin{pmatrix}\sqrt{Z_{\mathcal{H}_{1}}} & 0\\ 0 & \sqrt{Z_{\mathcal{H}_{2}}}\end{pmatrix} D(\alpha)^{-1} \begin{pmatrix} h \\ H \end{pmatrix}
  &&\equiv \mathbf{Z}_{hH} \begin{pmatrix}h\\ H \end{pmatrix},\\
  \begin{pmatrix}A\\ G \end{pmatrix} &\rightarrow D(\beta_{n})\begin{pmatrix}\sqrt{Z_{\mathcal{H}_{1}}} & 0\\ 0 & \sqrt{Z_{\mathcal{H}_{2}}}\end{pmatrix} D(\beta_{n})^{-1} \begin{pmatrix}A\\ G \end{pmatrix}
  &&\equiv \begin{pmatrix}Z_{AA} & Z_{AG} \\ Z_{GA} & Z_{GG} \end{pmatrix}  \begin{pmatrix}A\\ G\end{pmatrix} \; \equiv\; \mathbf{Z}_{AG}  \begin{pmatrix}A\\ G\end{pmatrix},\\
  \begin{pmatrix}H^{\pm}\\ G^{\pm}\end{pmatrix} &\rightarrow D(\beta_{c})\begin{pmatrix}\sqrt{Z_{\mathcal{H}_{1}}} & 0\\ 0 & \sqrt{Z_{\mathcal{H}_{2}}}\end{pmatrix} D(\beta_{c})^{-1} \begin{pmatrix}H^{\pm}\\ G^{\pm}\end{pmatrix}
  &&\equiv  \mathbf{Z}_{H^{\pm}G^{\pm}}  \begin{pmatrix}H^{\pm}\\  G^{\pm} \end{pmatrix} .
\end{alignat}
\end{subequations}
The matrices $\mathbf{Z}_{ij}$ can be expanded as
\begin{align}
\label{eq:fieldrenconstants}
\mathbf{Z}_{ij} =
\mathbf{1} + \delta^{(1)}\mathbf{Z}_{ij} +
\delta^{(2)}\mathbf{Z}_{ij}\, . 
\end{align}
The required one-loop expressions for the entries in the
$\delta^{(1)}\mathbf{Z}_{ij}$-matrices are given in
Ref.~\cite{Frank:2006yh}; the corresponding set of two-loop
expressions is given in Ref.~\cite{Hollik:2014bua}.

\medskip

The genuine two-loop counterterms $\delta^{(2)} \MZ_{hHA}$ of
Eq.~(\ref{eq:renselfenergies}) can now be summarized as
\begin{align}
\label{eq:higgs2Lct}
  \begin{split}
    \delta^{(2)}\MZ_{hHA} &= \delta^{(2)}\mathbf{M}_{hHA} +
      \begin{pmatrix} \delta^{(2)}\mathbf{Z}_{hH}^T & \mathbf{0} \\
                                \mathbf{0}  & \delta^{(2)}Z_{AA} \end{pmatrix} 
     \left(\mathbf{M}_{hHA}^{(0)} - p^2 \unity\right) \\
     &\quad+ \left(\mathbf{M}_{hHA}^{(0)} - p^2 \unity\right)  
     \begin{pmatrix} \delta^{(2)}\mathbf{Z}_{hH} & \mathbf{0} \\ \mathbf{0}  & \delta^{(2)}Z_{AA} \end{pmatrix}\ ,
  \end{split}
\end{align}
where the required two-loop mass counterterms read
{\allowdisplaybreaks
\begin{subequations}\label{eq:atatMassCT_2L}
\begin{align}
  \begin{split}
    \delta^{(2)}m_{h}^{2} &= c_{\alpha-\beta}^{2}\,\delta^{(2)}m_A^{2} + s_{\alpha+\beta}^{2}\,\delta^{(2)}m_Z^{2} + c_{\beta}^{2}\,\delta^{(2)}t_{\beta} \left(s_{2(\alpha-\beta)}\,m_A^{2} + s_{2(\alpha+\beta)}\,m_Z^{2}\right)\\
                       &\quad + \frac{e\,s_{\alpha-\beta}}{2\,M_{W}\,s_{\text{w}}} \Biggl[\left(1 + c_{\alpha-\beta}^{2}\right) \delta^{(2)}T_{h} + s_{\alpha-\beta}\,c_{\alpha-\beta}\,\delta^{(2)}T_{H}\Biggr]\ ,
  \end{split}\\[2ex]
  \begin{split}
    \delta^{(2)}m_{H}^{2} &= s_{\alpha-\beta}^{2}\,\delta^{(2)}m_A^{2} + c_{\alpha+\beta}^{2}\,\delta^{(2)}m_Z^{2} - c_{\beta}^{2}\,\delta^{(2)}t_{\beta} \left(s_{2(\alpha-\beta)}\,m_A^{2} + s_{2(\alpha+\beta)}\,m_Z^{2}\right)\\
                           &\quad - \frac{e\,c_{\alpha-\beta}}{2\,M_{W}\,s_{\text{w}}} \Biggl[\left(1 + s_{\alpha-\beta}^{2}\right) \delta^{(2)}T_{H} + c_{\alpha-\beta}\,s_{\alpha-\beta}\,\delta^{(2)}T_{h}\Biggr]\ ,
  \end{split}\\[2ex]
  \delta^{(2)}m_{A}^{2} &= \delta^{(2)}m_{H^\pm}^{2} - \delta^{(2)}m_{W}^{2}\ ,\\[2ex]
  \begin{split}
    \delta^{(2)}m_{hH}^{2} &= \frac{1}{2}\left(s_{2\left(\alpha-\beta\right)}\,\delta^{(2)}m_A^{2} - s_{2\left(\alpha+\beta\right)}\,\delta^{(2)}m_Z^{2}\right) - c_{\beta}^{2}\,\delta^{(2)}t_{\beta} \left(c_{2(\alpha-\beta)}\,m_A^{2} + c_{2(\alpha+\beta)}\,m_Z^{2}\right)\\
                           &\quad + \frac{e}{2\,M_{W}\,s_{\text{w}}} \Biggl[s_{\alpha-\beta}^{3}\,\delta^{(2)}T_{H} - c_{\alpha-\beta}^{3}\,\delta^{(2)}T_{h}\Biggr]\ ,
  \end{split}\\[2ex]
  \delta^{(2)}m_{hA}^{2} &= \frac{e}{2\,M_{W}\,s_{\text{w}}}\,s_{\alpha-\beta}\,\delta^{(2)}T_{A}\ ,\\[2ex]
  \delta^{(2)}m_{HA}^{2} &= -\frac{e}{2\,M_{W}\,s_{\text{w}}}\,c_{\alpha-\beta}\,\delta^{(2)}T_{A}\ .
\end{align}
\end{subequations}
}%
The entries of $\delta^{(2)}\mathbf{M}_{hHA}$ that are not listed here
are determined by symmetry. When replacing~$\delta^{(2)} \to \delta$
they are formally equal to the one-loop counterterms listed in
Eqs.~(53) of Ref.~\cite{Frank:2006yh} (up to the different convention
for the counterterm of~$t_\beta$ used there).

\medskip
The two-loop renormalization constants of
Eqs.~\eqref{eq:tadpolct}--\eqref{eq:atatMassCT_2L} are fixed by
extending the renormalization scheme of Ref.~\cite{Frank:2006yh} from
the one-loop to the two-loop order:
\begin{itemize}
\item 
The tadpole counterterms $\delta^{(2)} T_i $ are fixed by requiring
that the minimum of the Higgs potential is not shifted, which means
that the tadpole coefficients have to vanish at each order. At the
two-loop level, the condition reads
\begin{align}
  T_{i}^{(2)} + \delta^{(2)}T_{i}  &= 0\ , \quad i = h,\,H,\,A\ , 
\end{align}
where the $T_{i}^{(2)}$ denote the unrenormalized one-point functions
at two-loop order, see Fig.~\ref{fig:tadpoles} for the contributing
two-loop diagrams.  The aforementioned relation for the mixing
angles \mbox{$\beta_{n} = \beta_{c} = \beta$} is a consequence of the
tadpole conditions \mbox{$T_i = 0$} at lowest order.

\begin{figure}[tp!]
  \centering
  \includegraphics[width=.85\linewidth]{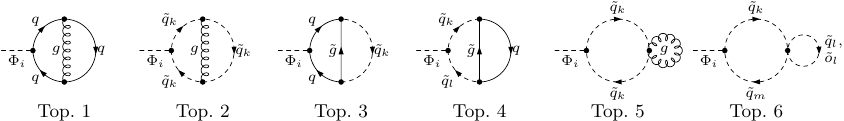}\\[-2ex]
  \rule{.85\linewidth}{\arrayrulewidth}
  \includegraphics[width=.567\linewidth]{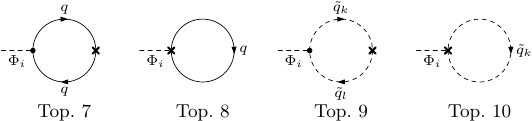}
  \caption{\label{fig:tadpoles} Types of two-loop tadpole diagrams
    contributing to $T^{(2)}_i$.  One-loop counterterm insertions are
    denoted by a cross.  \mbox{$\Phi_{i} = h,\,H,\,A$};
    \mbox{$\tilde{q} \neq \tilde{o}$}. Topology~$5$ contains a
    one-point loop with a mass-less gluon and is therefore equal to
    zero.}

  \capstart
  \vspace{6ex}

  \includegraphics[width=.85\linewidth]{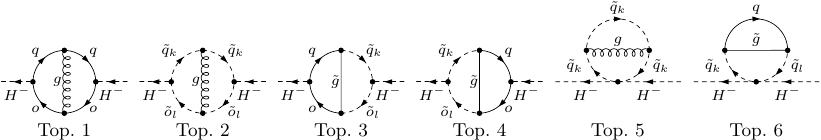}\\[-2ex]
  \rule{.85\linewidth}{\arrayrulewidth}
  \includegraphics[width=.85\linewidth]{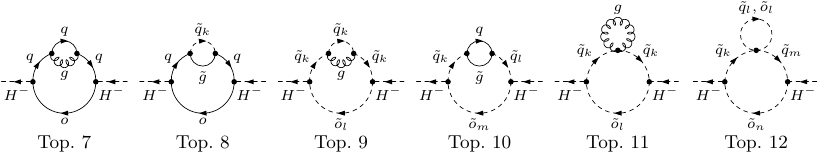}\\[-2ex]
  \rule{.85\linewidth}{\arrayrulewidth}
  \includegraphics[width=.425\linewidth]{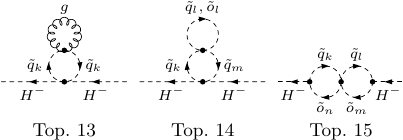}\\[-2ex]
  \rule{.85\linewidth}{\arrayrulewidth}
  \includegraphics[width=.85\linewidth]{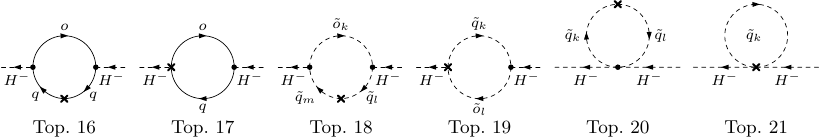}
  \caption{\label{fig:selfenergiescharged} Types of two-loop
    self-energy diagrams for the charged Higgs bosons.  One-loop
    counterterm insertions are denoted by a cross.  \mbox{$\;q \neq
      o$}, \mbox{$\;\tilde{q} \neq
      \tilde{o}$}. Topologies~$11$ and~$13$ contain a one-point
    loop with a mass-less gluon and are therefore equal to zero.}

  \capstart
  \vspace{6ex}

  \includegraphics[width=0.85\textwidth]{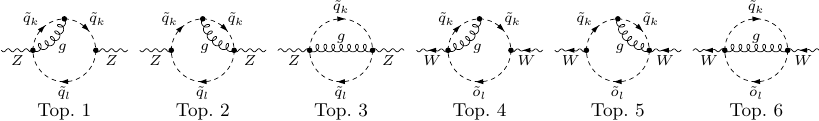} \caption{\label{fig:selfenergiesGauge}
  Additional types of two-loop self-energy diagrams for the gauge
  bosons besides the ones in analogy to
  Fig.~\ref{fig:selfenergiesneutral}
  and~\ref{fig:selfenergiescharged}.  \mbox{\mbox{$\;\tilde{q} \neq \tilde{o}$}.
  }}
\end{figure}

\item The charged Higgs-boson mass $m_{H^\pm}$ is the only independent
  mass parameter of the Higgs sector and is used as an input
  quantity. Accordingly, the corresponding mass counterterm is fixed
  by an independent renormalization condition, chosen as on-shell,
  given by
\begin{align}
  \Real{\hat{\Sigma}_{H^{\pm}}^{(2)}(m_{H^{\pm}}^2)} = 0 \text{ .}
\end{align}
The renormalized charged-Higgs self-energy at the two-loop level can
be expressed in terms of the unrenormalized charged self-energy and
its respective counterterms
\begin{align}
  \label{eq:renchargedselfenergy} 
  \hat{\Sigma}_{H^\pm}^{(2)} (m_{H^{\pm}}) = \Sigma_{H^\pm}^{(2)} (m_{H^{\pm}}^2) - \delta^{(2)}m^2_{H^{\pm}}\, ,
\end{align}
leading to the mass counterterm
\begin{align}
  \delta^{(2)}m_{H^{\pm}}^{2} &= \Real{\Sigma_{H^\pm}^{(2)}{\left(m_{H^{\pm}}^2\right)}}
\end{align}
when applying the on-shell condition.  The contributing Feynman
diagrams are shown in Fig.~\ref{fig:selfenergiescharged}. As we
neglect flavor mixing~$q$, $o$, $\tilde{q}$ and~$\tilde{o}$ always
belong to the same generation. As a consequence, the vertices with
four squarks in topologies~$12$, $14$ and~$15$ are only non-zero when
all adjacent fields are of the same generation.

\item The field-renormalization constants of the Higgs mass
  eigenstates in Eq.~\eqref{eq:neutralhiggsfieldren} are combinations
  of the basic doublet-field renormalization constants
  $\delta^{(2)}Z_{\mathcal{H}_{1}}$ and
  $\delta^{(2)}Z_{\mathcal{H}_{2}}$, which are fixed by the
  UV-divergent parts of the derivatives of the corresponding
  self-energies,
\begin{alignat}{4}
  \delta^{(2)}Z_{\mathcal{H}_{1}} &= 
  -\left[ \frac{d\Sigma^{(2)}_{HH}(p^2)}{dp^2} \right]_{\alpha=0}^{\text{div}} , &\quad
  \delta^{(2)}Z_{\mathcal{H}_{2}} &=
  -\left[ \frac{d\Sigma^{(2)}_{hh}(p^2)}{dp^2} \right]_{\alpha=0}^{\text{div}} .
\end{alignat}

\item Also $t_\beta$ is renormalized by a purely
  UV-divergent counterterm, which was shown to be a convenient
  choice~\cite{Freitas:2002um} (see also
  Refs.~\cite{Sperling:2013eva,Sperling:2013xqa}). Alternative
  process-dependent definitions for the renormalization of~$t_{\beta}$
  can be found in Ref.~\cite{Baro:2008bg}. For the class of two-loop
  corrections of~$\mathcal{O}{\left(\alpha \alpha_s\right)}$ the
  counterterm can be written as \begin{align} \delta^{(2)}t_{\beta}^2
  &= t_{\beta}^2 \left(\delta^{(2)}Z_{\mathcal{H}_{2}}
  - \delta^{(2)}Z_{\mathcal{H}_{1}}\right).  \end{align}

\item When neglecting momentum-dependent contributions and taking the
  gaugeless limit, the purely UV-divergent two-loop counterterms
  $\delta^{(2)}Z_{\mathcal{H}_{1}}$, $\delta^{(2)}Z_{\mathcal{H}_{2}}$
  and $\delta^{(2)}t_{\beta}$ cancel each other and are therefore not
  required for renormalization, compare
  Ref.~\cite{Passehr:2017ufr}. If one of these two limitations is
  dropped, $\delta^{(2)}Z_{\mathcal{H}_{1}}$,
  $\delta^{(2)}Z_{\mathcal{H}_{2}}$ and $\delta^{(2)}t_{\beta}$ are
  necessary in order to obtain a UV-finite result. In the corrections
  discussed in this article these counterterms have to be taken into
  account as none of these approximations is used.

  It should also be noted that the chosen renormalization conditions
  for $\delta^{(2)}Z_{\mathcal{H}_{2}}$ and $\delta^{(2)}t_{\beta}$
  are not equal to pure $\overline{\text{DR}}$ conditions, since the
  top-mass counterterm $\delta^{(1)}m_t$ which enters in
  $\delta^{(2)}Z_{\mathcal{H}_{2}}$ is fixed by an on-shell
  condition. The resulting differences between the two schemes have
  been discussed in~\cite{Degrassi:2014pfa,Borowka:2015ura}.

\item Renormalization of the $D$~terms in the Higgs--squark couplings
  which are induced by the gauge coupling~$g_2$, as well as the
  relation between the charged and $CP$-odd Higgs masses require
  counterterms for the $Z$- and $W$-boson masses, $\delta^{(2)}M_Z^2$
  and~$\delta^{(2)}M_W^2$, respectively. We treat~$M_W$ and~$M_Z$
  as independent input parameters and fix their renormalization
  constants by the on-shell conditions
  \begin{align}
  \Real{\hat{\Sigma}_{Z,W}^{(2)}(M_{Z,W}^2)} &= 0 \,,
  \end{align}
  leading to
  \begin{align}
  \delta^{(2)}M_{Z,W}^{2} &= \Real{\Sigma_{Z,W}^{(2)}{\left(M_{Z,W}^2\right)}}.
  \end{align}
  Here $\Sigma_{Z,W}^{(2)}$ denote the transverse parts of the two-loop
  self-energies of $Z$ and $W$, repectively.

  Most of the Feynman diagrams contributing to the two-loop
  self-energies~$\Sigma_{Z,W}^{(2)}$ differ from the Higgs
  self-energies depicted in Figs.~\ref{fig:selfenergiesneutral}
  and~\ref{fig:selfenergiescharged} only by the external
  fields. Pictorially, they can be otained by replacing the neutral
  external Higgs fields by the $Z$, and the charged Higgs field by the
  $W$ boson field. All additional topologies are depicted in
  Fig.~\ref{fig:selfenergiesGauge}.
\end{itemize}

\tocsubsection[\label{sec:subren}]{Sub-loop renormalization}

Apart from the genuine two-loop diagrams, the lowest-order QCD
contributions to the self-energies and tadpoles involve one-loop
diagrams with insertions of one-loop counterterms.  This
subrenormalization concerns masses and mixing of the colored
particles.

The required one-loop counterterms for subrenormalization arise from
the quark~$q$ and scalar quark~$\tilde{q}$ sectors.  The squark mass
matrices in the
$\big(\tilde{q}_{\text{L}},\,\tilde{q}_{\text{R}}\big)$ bases are
given in lowest order by
\begin{align}
  \label{eq:squarks}
    \mathbf{M}_{\tilde{q}} &= 
    \begin{pmatrix}
     m_{\tilde{q}_{\text{L}}}^{2} + m_{q}^{2} + M_Z^2\,c_{2\beta} (T_q^3 - Q_q\,s^2_{\mathrm{w}}) & 
     m_{q}\left(A_{q}^{*} - \mu\,\kappa_q \right)\\[0.1cm]
     m_{q}\left(A_{q} - \mu^{*}\,\kappa_q \right) & 
     m_{\tilde{q}_{\text{R}}}^{2} + m_{q}^{2} + M_Z^2\,c_{2\beta}\,Q_q\,s^2_{\mathrm{w}}
   \end{pmatrix}, & \begin{split}
     \kappa_{t,c,u} &= \frac{1}{t_{\beta}}\,,\\ \kappa_{b,s,d} &= t_{\beta}\,,
    \end{split}
\end{align}
with $Q_q$ and $T^3_q$ denoting charge and isospin of $q \in \{u, c,
t, d, s, b\}$. For the sake of convenience we suppress repeating the
indices of the first and second generation in the following since
renormalization is analogous to the third
generation. $SU(2)$-invariance requires
\mbox{$m_{\tilde{t}_{\text{L}}}^{2} = m_{\tilde{b}_{\text{L}}}^{2}
  \equiv m_{\tilde{Q}_{3}}^{2}$}.

\noindent
The squark mass eigenvalues can be obtained from unitary
transformations,
\begin{align}
  \label{eq:squarkdiag}
  \mathbf{U}^{\tilde{q}}\mathbf{M}_{\tilde{q}}\mathbf{U}^{\tilde{q}\dagger}  &= 
  \mathrm{diag}{\left(m_{\tilde{q}_{1}}^{2},\, m_{\tilde{q}_{2}}^{2}\right)}.
\end{align}
Since $A_{q}$ and $\mu$ are complex parameters, the unitary matrices
$\mathbf{U}^{\tilde{q}}$ can be described by the mixing angle
$\theta_{\tilde{q}}$ and an additional phase $\varphi_{\tilde{q}}$.

The independent parameters which enter the two-loop calculation
through the quark--squark sector are: the quark masses~$m_q$, the soft
SUSY-breaking parameters~$m_{\tilde{Q}_{i}}$
and~$m_{\tilde{q}_{\text{R}}}$, and the complex trilinear
couplings~$A_{q} = \lvert A_{q}\rvert\,\E^{i\phi_{A_{q}}}$.  These
parameters have to be renormalized at the one-loop level,
\begin{align}
  m_q &\rightarrow m_q + \delta^{(1)}m_q\,,&
  m_{\tilde{q}_\text{L,R}}^2 &\rightarrow m_{\tilde{q}_\text{L,R}}^2 + \delta^{(1)}m_{\tilde{q}_\text{L,R}}^2\,,&
  A_q &\rightarrow A_q + \delta^{(1)}A_q\,,
\end{align}
thus defining transformations~$\mathbf{M}_{\tilde{q}}\to
\mathbf{M}_{\tilde{q}} + \delta^{(1)}\mathbf{M}_{\tilde{q}}$ for the
mass matrices in Eq.~\eqref{eq:squarks}.  The other free
parameter~$\mu$, which is related to the Higgsino sector, enters the
self-energies as well. However, the renormalization of $\mu$ does not
receive one-loop corrections of $\mathcal{O}{\left(\alpha_s\right)}$
and is therefore not part of the contributions considered in this
calculation.

\medskip

The individual renormalization conditions for the colored sector are
formulated as follows:
\begin{itemize}
\item Renormalization of the top quark mass is carried out in the
  on-shell scheme, \IE{}
  \begin{align}
  \label{eq:mtopren}
    \delta^{(1)}m_{t} &=
    m_{t} \, \Real{\frac{1}{2}\left(\Sigma_{t}^{\text{L}(1)}{\left(m_{t}^{2}\right)} 
        + \Sigma_{t}^{\text{R}(1)}{\left(m_{t}^{2}\right)}\right) + \Sigma_{t}^{\text{S}(1)}{\left(m_{t}^{2}\right)}},
  \end{align}
  where the quark self-energy is given in terms of its Lorentz
  decomposition
  \begin{align}
    \label{eq:Lorentz} 
    \Sigma_q (p) & =\,  \not{\! p}\, \omega_-\,  \Sigma_q^{\mathrm{L}} (p^2) +
    \not{\! p}\, \omega_+\,   \Sigma_q^{\mathrm{R}}(p^2) 
    + m_q \,\Sigma_q^{\mathrm{S}}(p^2)  
    + m_q \gamma_5\, \Sigma_q^{\mathrm{PS}}(p^2)
  \end{align}
  with the left-, right-handed
  projectors~$\omega_{-,+}=\tfrac{1}{2}\left(1\mp\gamma_5\right)$.

  The bottom mass is renormalized in the $\overline{\text{DR}}$ scheme
  (see Refs.~\cite{Dedes:2003km,Heinemeyer:2004xw,Heinemeyer:2010mm})
  at the scale $m_t^{\text{os}}$. The counterterm can be obtained by
  using the expression in analogy to the counterterm for the top quark
  mass in Eq.~(\ref{eq:mtopren}) and restricting to the UV-divergent
  contributions at the scale $m_t^{\text{os}}$.  The choice of a
  $\overline{\text{DR}}$ renormalization for $m_b$ is convenient in
  order to incorporate a resummation of~$\tan\beta$-enhanced
  contributions to the relation between the bottom quark mass and the
  bottom Yukawa coupling, see Sec.~\ref{sec:bottomresum} below.  The
  contributing Feynman diagrams for the renormalization of~$m_t$
  and~$m_b$ are depicted in Fig.~\ref{fig:RCTop}.

\item In order to fix the renormalization constants of the stop
  sector, we employ the
  relation \begin{align} \label{eq:stopcountermmatrix} \delta^{(1)}\mathbf{M}_{\tilde{t}}
  &= \delta^{(1)}{\left(\mathbf{U}^{\tilde{t}\dagger}\,\mathrm{diag}{\left(m_{\tilde{q}_{1}}^{2},\,
  m_{\tilde{q}_{2}}^{2}\right)}\,\mathbf{U}^{\tilde{t}}\right)}
  = \mathbf{U}^{\tilde{t}\dagger}\begin{pmatrix} \delta^{(1)}m_{\tilde{t}_{1}}^{2}
  & \delta^{(1)}m_{\tilde{t}_{1}\tilde{t}_{2}}^{2} \\ \delta^{(1)}m_{\tilde{t}_{1}\tilde{t}_{2}}^{2\,*}
  & \delta^{(1)}m_{\tilde{t}_{2}}^{2} \end{pmatrix}\mathbf{U}^{\tilde{t}}\,.  \end{align}
  Thus we
  derive \begin{subequations} \begin{align}\label{eq:stopleft} \delta^{(1)}m_{\tilde{t}_{\text{L}}}^2
  &= \sum\limits_{i=1}^2\lvert\mathbf{U}^{\tilde{t}}_{i1}\rvert^{2}\,\delta^{(1)}m_{\tilde{t}_{i}}^{2}
  +
  2\,\Real{\mathbf{U}^{\tilde{t}}_{21}\mathbf{U}^{\tilde{t}*}_{11}\,\delta^{(1)}m_{\tilde{t}_{1}\tilde{t}_{2}}^{2}}
  -
  2\,m_{t}\,\delta^{(1)}m_{t}\,,\\ \delta^{(1)}m_{\tilde{t}_{\text{R}}}^2
  &= \sum\limits_{i=1}^2\lvert\mathbf{U}^{\tilde{t}}_{i2}\rvert^{2}\,\delta^{(1)}m_{\tilde{t}_{i}}^{2}
  +
  2\,\Real{\mathbf{U}^{\tilde{t}}_{22}\mathbf{U}^{\tilde{t}*}_{12}\,\delta^{(1)}m_{\tilde{t}_{1}\tilde{t}_{2}}^{2}}
  -
  2\,m_{t}\,\delta^{(1)}m_{t}\,,\\ \label{eq:Atrenormalization} \begin{split} \delta^{(1)}A_t
  &= \mathbf{U}^{\tilde{t}}_{11}\mathbf{U}^{\tilde{t}*}_{12}\frac{\delta^{(1)}m_{\tilde{t}_{1}}^{2}
  - \delta^{(1)}m_{\tilde{t}_{2}}^{2}}{m_t}
  + \mathbf{U}^{\tilde{t}}_{21}\mathbf{U}^{\tilde{t}*}_{12}\frac{\delta^{(1)}m_{\tilde{t}_{1}\tilde{t}_{2}}^{2}}{m_t}
  + \mathbf{U}^{\tilde{t}}_{22}\mathbf{U}^{\tilde{t}*}_{11}\frac{\delta^{(1)}m_{\tilde{t}_{1}\tilde{t}_{2}}^{2\,*}}{m_t}\\
  &\quad - \left(A_{t}
  - \frac{\mu^{*}}{t_\beta}\right)\frac{\delta^{(1)}m_t}{m_t}\,.  \end{split} \end{align} \end{subequations}
  The counterterm $\delta^{(1)}A_t$ given in
  Eq.~\eqref{eq:Atrenormalization} provides the renormalization of the
  complex parameter $A_t$. It should be noted that the imaginary parts
  of renormalization constants and parameters only appear in
  real-valued combinations in the Higgs self-energy.
  
  The counterterms~$\delta^{(1)}m_{\tilde{t}_{1}}^{2}$
  and~$\delta^{(1)}m_{\tilde{t}_{2}}^{2}$ are fixed by on-shell
  conditions for the top-squarks,
  \begin{align}
    \label{eq:stoponshell}
    \delta^{(1)}m_{\tilde{t}_{i}}^{2} &= \Real{\, \Sigma_{\tilde{t}_{ii}}^{(1)}{\left(m_{\tilde{t}_{i}}^{2}\right)}}, 
    \quad i=1,2\,,
  \end{align}
  involving the diagonal~$\tilde{t}_{1,2}$ self-energies, see
  Fig.~\ref{fig:RCTop}. The remaining
  counterterm~$\delta^{(1)}m_{\tilde{t}_{1}\tilde{t}_{2}}^{2}$ is
  fixed by the renormalization condition (see
  Ref.~\cite{Heinemeyer:2007aq})
  \begin{align}
    \label{eq:stopoffdiag}
    \delta^{(1)}m_{\tilde{t}_{1}\tilde{t}_{2}}^{2} &= 
    \frac{1}{2}\,\Real{\Sigma_{\tilde{t}_{12}}^{(1)}{\left(m_{\tilde{t}_{1}}^{2}\right)}
      + \Sigma_{\tilde{t}_{12}}^{(1)}{\left(m_{\tilde{t}_{2}}^{2}\right)}}\ ,
  \end{align}
  which involves the non-diagonal squark self-energy shown in
  Fig.~\ref{fig:RCTop} with incoming~$\tilde{t}_2$ and
  outgoing~$\tilde{t}_1$.

\begin{figure}[b!]
  \centering
  \vspace{2ex}
  \includegraphics[width=.85\textwidth]{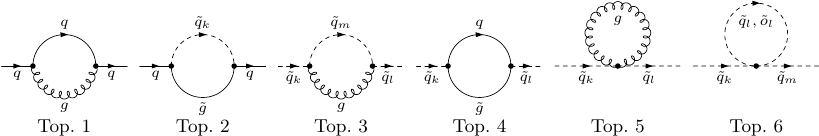}
  \caption{\label{fig:RCTop} Types of Feynman diagrams for the
  renormalization of the quark--squark
  sector.  \mbox{$\;\tilde{q} \neq \tilde{o}$}. Topology~$5$ contains
  a one-point loop with a mass-less gluon and is therefore equal to
  zero.}
\end{figure}
\clearpage
\item Between the gauge and mass
  eigenstates of the bottom squarks we employ an analogous relation to
  Eq.~\eqref{eq:stopcountermmatrix}. We derive 
  \begin{subequations}
  \begin{align}\label{eq:sbottomleft}
    \delta^{(1)}m_{\tilde{b}_{\text{L}}}^2 &= \sum\limits_{i=1}^2\lvert\mathbf{U}^{\tilde{b}}_{i1}\rvert^{2}\,\delta^{(1)}m_{\tilde{b}_{i}}^{2} + 2\,\Real{\mathbf{U}^{\tilde{b}}_{21}\mathbf{U}^{\tilde{b}*}_{11}\,\delta^{(1)}m_{\tilde{b}_{1}\tilde{b}_{2}}^{2}} - 2\,m_{b}\,\delta^{(1)}m_{b}\,,\\
    \delta^{(1)}m_{\tilde{b}_{\text{R}}}^2 &= \sum\limits_{i=1}^2\lvert\mathbf{U}^{\tilde{b}}_{i2}\rvert^{2}\,\delta^{(1)}m_{\tilde{b}_{i}}^{2} + 2\,\Real{\mathbf{U}^{\tilde{b}}_{22}\mathbf{U}^{\tilde{b}*}_{12}\,\delta^{(1)}m_{\tilde{b}_{1}\tilde{b}_{2}}^{2}} - 2\,m_{b}\,\delta^{(1)}m_{b}\,,\\
    \label{eq:Abrenormalization}
    \begin{split}
      \delta^{(1)}A_b &= \Bigg[\mathbf{U}^{\tilde{b}}_{11}\mathbf{U}^{\tilde{b}*}_{12}\frac{\delta^{(1)}m_{\tilde{b}_{1}}^{2} - \delta^{(1)}m_{\tilde{b}_{2}}^{2}}{m_b}
      + \mathbf{U}^{\tilde{b}}_{21}\mathbf{U}^{\tilde{b}*}_{12}\frac{\delta^{(1)}m_{\tilde{b}_{1}\tilde{b}_{2}}^{2}}{m_b} + \mathbf{U}^{\tilde{b}}_{22}\mathbf{U}^{\tilde{b}*}_{11}\frac{\delta^{(1)}m_{\tilde{b}_{1}\tilde{b}_{2}}^{2\,*}}{m_b}\\
      &\quad\hphantom{\Bigg[} - \left(A_{b} - \mu^{*}\,t_\beta\right)\frac{\delta^{(1)}m_b}{m_b}\Bigg]_{\overline{\text{DR}}}\,.
    \end{split}
  \end{align}
  \end{subequations}
  As indicated by the subscript, we choose to renormalize~$A_b$ in
  the~$\overline{\text{DR}}$~scheme, which has been shown to be
  convenient for reasons of numerical
  stability~\cite{Dedes:2003km,Heinemeyer:2004xw,Heinemeyer:2010mm}.
  The scale of~$A_b$ is chosen to be~$m_t^{\text{os}}$.

  As a consequence of~$SU(2)$~invariance
  the counterterm~$\delta^{(1)}m_{\tilde{b}_{\text{L}}}^{2}$ is not
  independent, but a derived quantity which is fixed by the
  renormalization of the top--stop sector in Eq.~\eqref{eq:stopleft},
  since
  \begin{gather}
    \delta^{(1)}m_{\tilde{b}_{\text{L}}}^{2} = \delta^{(1)}m_{\tilde{Q}_{3}}^{2} = \delta^{(1)}m_{\tilde{t}_{\text{L}}}^{2}\,.
  \end{gather}
  Inserting Eq.~\eqref{eq:sbottomleft} and solving
  for~$\delta^{(1)}m_{\tilde{b}_1}^{2}$ yields
  \begin{align}
    \label{eq:sbottomdependent}
    \delta^{(1)}m_{\tilde{b}_{1}}^{2} &= \frac{1}{\lvert\mathbf{U}^{\tilde{b}}_{11}\rvert^{2}}\left(\delta^{(1)}m_{\tilde{t}_{\text{L}}}^{2} - \lvert\mathbf{U}^{\tilde{b}}_{12}\rvert^{2}\,\delta^{(1)}m_{\tilde{b}_{2}}^{2} - 2\,\Real{\mathbf{U}^{\tilde{b}}_{21}\mathbf{U}^{\tilde{b}*}_{11}\,\delta^{(1)}m_{\tilde{b}_{1}\tilde{b}_{2}}^{2}} - 2\,m_{b}\,\delta^{(1)}m_{b}\right).
  \end{align}
  The other two counterterms~$\delta^{(1)}m_{\tilde{b}_{2}}^{2}$
  and~$\delta^{(1)}m_{\tilde{b}_{1}\tilde{b}_{2}}^{2}$ are fixed
  analogously as for the stops:
  \begin{subequations}
  \begin{align}
    \label{eq:sbottomonshell}
    \delta^{(1)}m_{\tilde{b}_2}^{2} &= \Real{\, \Sigma_{\tilde{b}_{22}}^{(1)}{\left(m_{\tilde{b}_2}^{2}\right)}}\,,\\
    \delta^{(1)}m_{\tilde{b}_{1}\tilde{b}_{2}}^{2} &= 
    \frac{1}{2}\,\Real{\Sigma_{\tilde{b}_{12}}^{(1)}{\left(m_{\tilde{b}_{1}}^{2}\right)}
      + \Sigma_{\tilde{b}_{12}}^{(1)}{\left(m_{\tilde{b}_{2}}^{2}\right)}}\,.
  \end{align}
  \end{subequations}
  Therefore in our scheme 
  only~$m_{\tilde{b}_2}$ is renormalized
  on-shell, while the counterterm $\delta^{(1)}m_{\tilde{b}_1}^{2}$ is a
  derived quantity according to Eq.~(\ref{eq:sbottomdependent}).
\end{itemize}

\tocsubsection[\label{sec:bottomresum}]{Resummation of \texorpdfstring{$\tan\beta$}{tan \unicodebeta}-enhanced terms}
\label{sec:botresum}

The Yukawa coupling of the bottom quark~$h_b$ receives radiative
corrections proportional to $\tan\beta$. Those $\tan\beta$-enhanced
contributions can be resummed as described in
Refs.~\cite{Banks:1987iu,Hall:1993gn,Hempfling:1993kv,Carena:1994bv,Carena:1999py,Eberl:1999he,Williams:2011bu}. The
resummed contributions $\Delta_b$ are UV finite and generally yield
complex numerical results. For the numerical evaluation in
Sec.~\ref{sec:numeric}, we use the version for $\Delta_b$ at the
one-loop order which is implemented in \texttt{FeynHiggs} and outlined
in the following. The largest~$\tan\beta$-enhanced contributions can
be absorbed by using an effective bottom-quark mass, which is related
to the~$\overline{\text{DR}}$-renormalized bottom quark mass in the
MSSM as follows,
\begin{align}\label{eq:mbeff}
  m_b^{\overline{\text{DR}},\text{MSSM}}{\left(m_t^{\text{os}}\right)} \simeq m_{b,\text{eff}} &= \frac{m_b^{\overline{\text{DR}},\text{SM}}{\left(m_t^{\text{os}}\right)}}{\lvert 1 + \Delta_b \rvert}\left(1 - \delta_b\right),
\end{align}
where $m_b^{\overline{\text{DR}},\text{SM}}(m_t^{\text{os}})$ is the
bottom mass in the $\overline{\text{DR}}$ renormalization scheme in
the Standard Model evaluated at the on-shell top mass. The
$\tan\beta$-enhanced contributions are captured in $\Delta_b$ and
properly resummed by including them in the denominator. The remaining
parts of the scalar part of the $\overline{\text{DR}}$-renormalized
bottom self-energy~$\hat{\Sigma}_b^{\text{S}}$ which are not enhanced
by~$\tan\beta$ are contained in~$\delta_b$ such that
\begin{align}
  \hat{\Sigma}_b^{\text{S}}{\left(0\right)} &= -\Delta_b - \delta_b\,.
\end{align}
The expression $\Delta_b$ at the one-loop order contains contributions
from gluinos, charginos and neutralinos (ordered in decreasing
numerical importance) and reads
\begin{align}
  \begin{split}
    \Delta_b &= \frac{2\,\alpha_s{\left(Q\right)}}{3\pi}\frac{M_3^*}{m_b}\sum\limits_{i=1}^2\mathbf{U}^{\tilde{b}}_{i1}\mathbf{U}^{\tilde{b}*}_{i2}\,\Bnull{0,\lvert M_3\rvert^2,m_{\tilde{b}_i}^2}\\
    &\quad+\frac{\alpha{\left(Q\right)}}{4\pi}\sum\limits_{g=1}^3\sum\limits_{i,j=1}^2\frac{m_{\tilde{\chi}^\pm_i}}{m_b}\, c_{\text{L}}\, c_{\text{R}}\, \lvert\mathbf{C}_{g3}\rvert^2\,\Bnull{0,m_{\tilde{\chi}^\pm_i}^2,m_{\tilde{u}^g_j}^2}\\
    &\quad -\frac{\alpha{\left(Q\right)}}{8\pi}\sum\limits_{i=1}^4\sum\limits_{j=1}^2 \frac{m_{\tilde{\chi}^0_i}}{m_b}\, n_{\text{L}}\, n_{\text{R}}\, \Bnull{0,m_{\tilde{\chi}^0_i}^2,m_{\tilde{b}_j}^2}\ .
  \end{split}
\end{align}
The couplings $\alpha_s$ and $\alpha$ are running parameters and are
evaluated at the scale $Q =
\sqrt{\vphantom{Q}m_{\tilde{b}_1}m_{\tilde{b}_2}}$. The symbol
$\mathbf{C}$ depicts the~CKM~matrix, and $u^g,\,\tilde{u}^g$ are the
$g$th generation up-type quarks and squarks, whereas $\Bnull{0, m_1,
m_2}$ and $\Beins{0, m_1, m_2}$ are one-loop functions.\footnote{These
one-loop functions are given by
\begin{align*}
  \Beins{0, m_1, m_2} &= -\frac{1}{2} \Bnull{0, m_1, m_2} +
  \frac{m_2 - m_1}{2} D\!B_0{\left(0, m_1, m_2\right)}\ , &
  \Bnull{0, m_1, m_2} &= \frac{A_0{\left(m_1\right)} - A_0{\left(m_2\right)}}{m_1 - m_2}\ ,\\
  D\!B_0{\left(0, m_1, m_2\right)} &= \frac{m_1^2 - m_2^2 + 2\, m_1\,
    m_2 \log{\tfrac{m_2}{m_1}}}{2\, (m_1 - m_2)^3}\ , &
  A_0{\left(m\right)} &= m \left(1 -
    \log{\tfrac{m}{\mu_{r}}}\right)\ .
\end{align*}
}\ As mentioned above, we otherwise neglect CKM~mixing in the two-loop
contributions that we evaluate. The renormalization scale~$\mu_{r}$
from the loop integrals drops out of~$\Delta_b$. The coefficients
$c_{\text{L,R}}$ and~$n_{\text{L,R}}$ are given by
\begin{subequations}
\begin{align}
  c_{\text{L}} &= \frac{\mathbf{V}^{\tilde{\chi}*}_{i1}\,\mathbf{U}^{\tilde{u}^g}_{j1}}{s_{\text{w}}} - \frac{m_{u^g}\,\mathbf{V}^{\tilde{\chi}*}_{i2}\,\mathbf{U}^{\tilde{u}^g}_{j2}}{\sqrt{2}\,M_W\,s_\beta\,s_{\text{w}}}\,, &
  c_{\text{R}} &= \frac{m_b\,\mathbf{U}^{\tilde{\chi}*}_{i2}\,\mathbf{U}^{\tilde{u}^g*}_{j1}}{\sqrt{2}\,M_W\,c_\beta\,s_{\text{w}}}\,,\\[1ex]
  n_{\text{L}} &= \left(\frac{\mathbf{N}^{\tilde{\chi}*}_{i1}}{3\,c_{\text{w}}} - \frac{\mathbf{N}^{\tilde{\chi}*}_{i2}}{s_{\text{w}}}\right)\mathbf{U}^{\tilde{b}}_{j1} +
    \frac{m_b\,\mathbf{N}^{\tilde{\chi}*}_{i3}\,\mathbf{U}^{\tilde{b}}_{j2}}{M_W\,c_\beta\,s_{\text{w}}}\,, &
  n_{\text{R}} &= \frac{2\,\mathbf{N}^{\tilde{\chi}*}_{i1}\,\mathbf{U}^{\tilde{b}*}_{j2}}{3\,c_{\text{w}}} + \frac{m_b\,\mathbf{N}^{\tilde{\chi}*}_{i3}\,\mathbf{U}^{\tilde{b}*}_{j1}}{M_W\,c_\beta\,s_{\text{w}}}\,.
\end{align}
\end{subequations}
In order to obtain a full conversion of the bottom mass between the
on-shell scheme and the $\overline{\text{DR}}$~scheme in
Eq.~\eqref{eq:mbeff}, those parts of the bottom self-energy which are
not enhanced by~$\tan\beta$ are included in~$\delta_b$ and
incorporated in the numerator of Eq.~\eqref{eq:mbeff}. Here, we
set~\mbox{$\mu_r=m_t^{\text{os}}$}.

At the one-loop order they read
\begin{align}
  \begin{split}
    \delta b &= \frac{\alpha_s{\left(Q\right)}}{3\pi}\sum\limits_{i=1}^2\Beins{0,\lvert M_3\rvert^2,m_{\tilde{b}_i}^2}
    + \frac{\alpha{\left(Q\right)}}{8\pi}\sum\limits_{g=1}^3\sum\limits_{i,j=1}^2\left[\left|c_{\text{L}}\right|^2 + \left|c_{\text{R}}\right|^2\right]\lvert\mathbf{C}_{g3}\rvert^2\,\Beins{0,m_{\tilde{\chi}^\pm_i}^2,m_{\tilde{u}^g_j}^2}\\[-.5ex]
    &\quad + \frac{\alpha{\left(Q\right)}}{16\pi}\sum\limits_{i=1}^4\sum\limits_{j=1}^2\left[\left|n_{\text{L}}\right|^2 + \left|n_{\text{R}}\right|^2\right]\Beins{0,m_{\tilde{\chi}^0_i}^2,m_{\tilde{b}_j}^2}\ .
  \end{split}
\end{align}
The parameters entering in $\Delta_b$ and $\delta_b$ are computed in
the limit of large $\tan\beta$. The chargino and neutralino masses and
mixing matrices are then obtained as
\begin{align}
   \lim_{t_\beta\rightarrow \infty} \text{diag}{\left(m_{\tilde{\chi}^\pm_1},m_{\tilde{\chi}^\pm_2}\right)} &= 
\mathbf{U}^{\tilde{\chi}*}\mathbf{X}\mathbf{V}^{\tilde{\chi}\dagger}\,, &
   \lim_{t_\beta\rightarrow \infty} \text{diag}{\left(m_{\tilde{\chi}^0_1},m_{\tilde{\chi}^0_2},m_{\tilde{\chi}^0_3},m_{\tilde{\chi}^0_4}\right)} &= 
\mathbf{N}^{\tilde{\chi}*}\mathbf{Y}\mathbf{N}^{\tilde{\chi}\dagger}\,,
\end{align}
where we use
\begin{align}
  \mathbf{X} &= \lim_{t_\beta\rightarrow \infty} \mathbf{M}_{\chi^\pm} =  \begin{pmatrix} M_2 & \sqrt{2} M_W s_\beta\\ 0 & \mu\end{pmatrix}, &
  \mathbf{Y} &=  \lim_{t_\beta\rightarrow \infty} \mathbf{M}_{\chi^0} = \begin{pmatrix} M_1 & 0 & 0 & M_Z s_\beta s_{\text{w}}\\ 0 & M_2 & 0 & -M_Z s_\beta c_{\text{w}}\\ 0 & 0 & 0 & -\mu\\ M_Z s_\beta s_{\text{w}} & -M_Z s_\beta c_{\text{w}} & -\mu & 0\end{pmatrix}.
\end{align}
Thereby the matrices $\mathbf{U}^{\tilde{\chi}}$ and
$\mathbf{V}^{\tilde{\chi}}$ yield a singular value decomposition for
$\mathbf{X}$, and the matrix~$\mathbf{N}^{\tilde{\chi}}$ yields
Takagi's factorization~\cite{Takagi:1927} on~$\mathbf{Y}$.

The sbottom masses in this limit are computed from the matrix given in
Eq.~\eqref{eq:squarks} at $A_b = 0$. Since the bottom mass itself also
enters that matrix, the final solution for $m_{b,\text{eff}}$ is found
iteratively.

By using Eq.~\eqref{eq:mbeff} for the bottom mass in the one-loop
contributions to the Higgs masses, the leading higher-order
corrections to the Higgs masses from the bottom--sbottom sector are
generated. The contributions of the bottom--sbottom sector to the
two-loop self-energies presented in this article add further
subleading shifts. It should be noted that the expression given in
Eq.~\eqref{eq:mbeff}, which employs the $\overline{\text{DR}}$~scheme
in the MSSM, is chosen such that no double counting of the terms
contained in~$m_{b,\text{eff}}$ occurs at the two-loop level.

\tocsection[\label{sec:numericeval}]{Numerical evaluation of the self-energies}

\begin{figure}[t!]
  \centering
  \includegraphics[width=.85\linewidth]{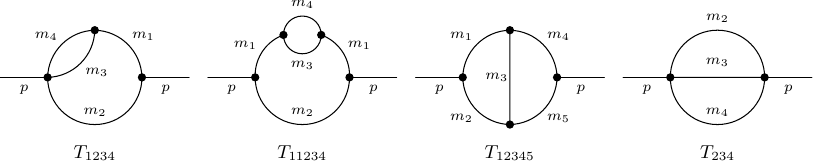}
  \caption{\label{fig:Tintegrals}Irreducible two-loop topologies resulting
  from tensor reduction, calculated numerically with the program
  \texttt{SecDec}. Some of the internal lines may also be massless.}
  \vspace{-2ex}
\end{figure}

The renormalized two-loop self-energies are expressed in terms of
two-loop two-point multi-scale integrals with non-zero external
momenta. With the help of~\texttt{TwoCalc}~\cite{Weiglein:1993hd}
and~\texttt{Reduze}~\cite{vonManteuffel:2012np} all integrals can be
reduced to the four irreducible scalar two-loop topologies depicted in
Fig.~\ref{fig:Tintegrals}, and products of analytically well-known
one-loop one- and two-point functions.

The scalar two-loop integrals are defined as
\begin{multline}
T_{i_1 i_2\dots i_n}(p^2, m_{i_1}^2,m_{i_2}^2,\dots, m_{i_5}^2) = \left(2 \pi \mu_r \right)^{2(4-D)}\\
\times \iint \frac{\text{d}^D  q_1}{i \pi^2}\, \frac{\text{d}^D  q_2}{i \pi^2}
\frac{1}{(k_{i_1}^2-m_{i_1}^2+ i \delta)(k_{i_2}^2-m_{i_2}^2+ i \delta)\cdots(k_{i_n}^2-m_{i_n}^2+ \I \delta)} \text{ ,}
\label{eq:tints}
\end{multline}
where $p$ is the external momentum, $q_i$ are the loop momenta, $m_i$
the masses of the propagators, $\mu_r$ is the renormalization scale
and $D=4-2\varepsilon$ the dimension. The $\I\delta$ results from the
solutions of the field equations in terms of causal Green functions,
while the indices ${i_1, i_2, \dots i_n}$ label which $k_{i}$ and
$m_i$ appear in the propagators of the integral. The five different
$k_{i}$ read
\begin{align}
\label{eq:kis}
k_1=q_1, \hspace{10pt} k_2=q_1+p, \hspace{10pt} k_3=q_2 -q_1, \hspace{10pt} k_4=q_2, \hspace{10pt} k_5=q_2+p \text{.}
\end{align}

The irreducible two-loop integrals of Fig.~\ref{fig:Tintegrals} may
depend on up to five different internal mass scales taken from the
following set,
\begin{align}
m_t,\, m_b,\, m_{\tilde{t}_1},\, m_{\tilde{t}_2},\, m_{\tilde{b}_1},\, m_{\tilde{b}_2},\, m_{\tilde{g}} = |M_3|,
\end{align}
in addition to a non-zero external momentum, taking the values of
$p^2=M_{h_1}^2,\, M_{h_2}^2,\, M_{h_3}^2$ when entering the
unrenormalized self-energies, or $p^2=m_{H^\pm}^2,\, m_W^2,\, m_Z^2$
when entering the self-energies through two-loop renormalization
constants. Recently, a lot of progress has been made towards
describing and evaluating integrals of this class analytically
\cite{Bloch:2013tra,Bloch:2014qca,Adams:2013kgc,Remiddi:2013joa,Adams:2015gva,Adams:2015ydq,Bloch:2016izu,Remiddi:2016gno,Adams:2016xah,Broedel:2017siw}.
However, to the best of our knowledge, an implementation of the
analytical results for all topologies in Fig.~\ref{fig:Tintegrals} is
not publicly available. We have therefore calculated these integrals
numerically using the
program~\texttt{SecDec}~\cite{Carter:2010hi,Borowka:2012yc,Borowka:2013cma}.

For the evaluation, the resulting new contributions to the neutral
Higgs-boson self-energies have been added to \texttt{FeynHiggs} via
its interface to external programs, see section~2.4 of
Ref.~\cite{Borowka:2014wla} for details. We have extended the existing
interface to the program~\texttt{SecDec} in~\texttt{FeynHiggs} to deal
with the 177 mass configurations of which 88 are computed at four
different kinematic points, 72 at two and the rest at one kinematic
point. The parameters entering the integrals are evaluated
by \texttt{FeynHiggs} and passed on to \texttt{SecDec}. It should be
noted that the heavy growth of mass configurations with respect to
non-electroweak corrections is due to an increase in the number of
mass scales involved in the renormalized self-energies.

We constructed two independent integration setups to allow for
consistency checks of the numerical result. The two-point one-loop
topologies entering the self-energies up to~$\mathcal{O}(\varepsilon)$
are known analytically. The bulk of their implementation was
previously tested in Ref.~\cite{Borowka:2014wla} and compared with the
authors of Ref.~\cite{Degrassi:2014pfa}. Additional mass
configurations were newly implemented and checked
against \texttt{SecDec}.  The increase in two-loop mass configurations
by more than a factor five with respect to the previous setup in
Ref.~\cite{Borowka:2014wla} calls for a higher precision of the
integrals to avoid numerical instabilites due to cancellations.  With
the integral reduction, unphysical thresholds can be introduced which
cancel in the sum of all contributing diagrams.  Numerically, due to
round-off errors, the cancellation might however not always be exact,
leading to numerical instabilities. The latter are cured by
introducing a small imaginary part to the denominators of the
coefficients arising from the integral reduction. We have verified
that the numerical dependence of the self-energies on this technical
regularization parameter is negligible.

The fact that we take a non-zero value of the bottom quark mass into
account leads to a large hierarchy among the different mass scales.
Numerical convergence at the desired accuracy is therefore difficult
to accomplish. On the other hand, we have analyzed the influence of
the quark masses of the first and second generation on the two-loop
integrals in the self-energies. For the second generation
and~$\tan{\beta}\gg 1$ a negative shift in the Higgs-boson mass
correction of only about $20$\,MeV can be observed when neglecting the
light quark masses. The effect is even smaller for the quark masses of
the first generation. The terms which involve the light quark masses
in couplings are negligible, too. It is due to this reason that we
will assume the first and second generation quarks to be massless
throughout the rest of our numerical analysis. The numerical impact of
the gauge contributions of the light quarks will be discussed below.

In order to achieve a relative precision of at least~$10^{-7}$ for
each integral, we use the deterministic integrator~\textsc{Cuhre}
included in the~\textsc{Cuba}
library~\cite{Agrawal:2011tm,Hahn:2004fe} but have optimized the
integration parameters for each integral topology and mass
configuration individually.

As a further crosscheck of our computation, we have compared
the~$\mathcal{O}{\left(\alpha\alpha_s\right)}$ contribution by the
top--stop and bottom--sbottom particles to the $Z$-boson self-energy
which is required for renormalization of the Higgs sector.
Since~\cite{Degrassi:2014pfa} uses massless bottom quarks, we have
reevaluated our result for the~$Z$ self-energy in the
limit~$m_b=0$. In order to avoid a dependence on the renormalization
scheme of the quark--squark sectors, the $Z$-boson self-energy has
been evaluated in the $\overline{\text{DR}}$~scheme by both groups for
this comparison. Overall we have found a very good agreement with
discrepancies at the level of~$0.3$\,GeV$^2$.

We find an overall uncertainty of the self-energies entering the light
Higgs-boson mass of maximally~$0.2\%$ by adding all uncertainties on
the numerical evaluation of the two-loop integrals in
quadrature. Given the resulting size of our newly computed corrections
analyzed in the next section, the absolute uncertainty on the light
Higgs boson mass is maximally~$0.4$\,MeV, which is well below the
shift coming from neglecting light quark masses from the first and
second generation.

The total of~$513$ integrals have been computed numerically on the fly
before passing the resulting two-loop self-energies back
to~\texttt{FeynHiggs}, where they are added to the corresponding
matrix elements just before the determination of the propagator poles.

\tocsection[\label{sec:numeric}]{Numerical results for the Higgs mass spectrum}

In the following we analyze the numerical impact of the newly computed
corrections. We start with a comparison with earlier results in the
literature and then discuss our results in three different scenarios:
an $m_h^{\text{mod}}$-like scenario (based on
Ref.~\cite{Carena:2013ytb}), a scenario with a particularly large
value of $\tan\beta$ where contributions from the bottom and sbottom
sector are enhanced, and a low-$m_H$ scenario (inspired by
Refs.~\cite{Carena:2013ytb,Bechtle:2016kui}).  For better readability
of the results, we define three different Higgs-boson masses resulting
from different higher-order contributions
\begin{alignat}{2}
  & M_{h_i}^{\text{old}}, && \text{ contains: } \mathcal{O}(\alpha_t\alpha_s)\rvert_{p^2=0} \text{ with complex parameters},\nonumber\\
  & \tilde{M}_{h_i}^{\text{old}}, && \text{ contains: } \text{same as } M_{h_i}^{\text{old}} + \left.\mathcal{O}(\alpha_b\alpha_s)\right|_{p^2=0} \text{ with real parameters},\nonumber\\
& M_{h_i}^{\text{new}}, && \text{ contains: } \mathcal{O}(\alpha_q\alpha_s),\mathcal{O}(\alpha\alpha_s),
  \mathcal{O}(h_{q}h_{o}\alpha_s)
 \text{ with non-zero } p^2,\nonumber\\
  \omit\span\omit\span\omit\span\text{with } i \in \{1,2,3\},\quad  q,o\in\{b,t\}. \label{eq:def_deltaMh}
\end{alignat}
All the above results contain the full one-loop and leading
$\mathcal{O}(\alpha_t^2)\rvert_{p^2=0}$ two-loop contributions, and
the $\tan\beta$-enhanced contributions to the relation between the
bottom quark mass and the bottom Yukawa coupling are resummed, see
section~\ref{sec:botresum}.  As mentioned earlier, the quark masses
and Yukawa couplings of the first and second family are
neglected. Thus, the first and second generation contributes only at
$\mathcal{O}(\alpha\alpha_s)$ by D-term contributions of the
sfermions. We focus our numerical discussion on the fixed-order result
up to the two-loop level, \IE~no combination with resummed
higher-order logarithmic contributions as discussed in
Refs.~\cite{Hahn:2013ria,Bahl:2016brp,Bahl:2017aev} is employed.

Using the definitions of Eq.~(\ref{eq:def_deltaMh}), we assign
\begin{align}
  \label{eq:deltamhdef}
  \Delta M_{h_i} = M_{h_i}^{\text{new}} - M_{h_i}^{\text{old}}, \quad
  \Delta \tilde{M}_{h_i} = M_{h_i}^{\text{new}} -
  \tilde{M}_{h_i}^{\text{old}}\,.
\end{align}
The size of the effects of our newly computed contributions is
contained in~$\Delta M_{h_i}$, since all the previously known terms
are subtracted. So far, the two-loop terms
of~$\mathcal{O}(\alpha_b\alpha_s)$ were only known in the MSSM with
real parameters and~$m_A$ as input parameter. $\Delta \tilde{M}_{h_i}$
shows our new contributions without these terms, if~$m_A$ is chosen as
input parameter.

Below we will discuss our results for non-zero phases of complex
parameters.  We investigate in particular the variation of the
phases~$\phi_{M_3},\,\phi_{A_t}$ and~$\phi_{A_b}$, which are much less
constrained by experimental bounds on EDMs than the phases of $\mu$,
$M_1$ (in the usual convention where the parameter $M_2$ is chosen to
be real) and the phases of the trilinear couplings of the first and
second generation.  As discussed \EG{} in Ref.~\cite{Arbey:2014msa},
scenarios with relatively large phase values are possible. In order to
demonstrate the possible impact of the phase variations on the Higgs
spectrum, below we display the phase dependences over the whole
range~$\left(-\pi,\pi\right]$.

\tocsubsection{Comparison with earlier results}

In a first step, in Tab.~\ref{tab:comparison} we show a comparison of
the results for the light Higgs-boson mass including our new
contributions with the results of Ref.~\cite{Degrassi:2014pfa}, where
in the MSSM with real parameters the corrections of
$\mathcal{O}(\alpha_t\alpha_s p^2)$ and the full corrections of
$\mathcal{O}(\alpha\alpha_s)$ have been evaluated, and with the
results up to $\mathcal{O}(\alpha_t\alpha_s p^2)$ in the MSSM with
real parameters from Ref.~\cite{Borowka:2014wla}. The comparison is
carried out for the benchmark scenarios $m_h^{\text{max}}$,
$m_h^{\text{mod}+}$, $m_h^{\text{mod}-}$ defined in
Ref.~\cite{Carena:2013ytb} and for a modified light-stop scenario used
in Ref.~\cite{Bagnaschi:2014zla}. We find overall good agreement with
the results of Ref.~\cite{Degrassi:2014pfa}. The comparison of the
corrections of $\mathcal{O}(\alpha_t\alpha_s p^2)$ with the full
corrections of $\mathcal{O}(\alpha\alpha_s)$ shows that the inclusion
of momentum dependence in the $\mathcal{O}(\alpha_t\alpha_s p^2)$
corrections yields a downward shift in $M_h$ which is to a large
extent compensated by the further corrections of
$\mathcal{O}(\alpha\alpha_s)$ for the scenarios that are considered
here.  The corrections beyond those of $\mathcal{O}(\alpha_t\alpha_s
p^2)$ yield an upward shift in $M_h$ of $520$\,MeV in the
$m_h^{\text{mod}+}$ and more than~$1$\,GeV in the $m_h^{\text{mod}-}$
scenario compared to the results of Ref.~\cite{Borowka:2014wla}.  The
size of the corrections shows a significant dependence on the
parameters in the stop sector.  The corrections are largest in the
$m_h^{\text{mod}-}$ scenario, where the stop masses are near the SUSY
scale and $A_t$ is negative. In this case there is a large
compensation between the downward shift caused by the corrections of
$\mathcal{O}(\alpha_t\alpha_s p^2)$ and the upward shift caused by the
further corrections of $\mathcal{O}(\alpha\alpha_s)$.  On the other
hand, the corrections are smallest for the modified light-stop
scenario, in which case we find that the contributions beyond the ones
of $\mathcal{O}(\alpha_t\alpha_s p^2)$ from
Ref.~\cite{Borowka:2014wla} even yield a small downward shift.  The
numerical differences between the results for the contributions of
$\mathcal{O}(\alpha_t\alpha_s p^2)$ from Ref.~\cite{Degrassi:2014pfa}
and Ref.~\cite{Borowka:2014wla}, which amount up to 0.3\,GeV for the
examples considered here, result from different renormalization scheme
choices of $\delta^{(2)}Z_{\mathcal{H}_{i}}$, see the discussion in
Refs.~\cite{Borowka:2014wla,Degrassi:2014pfa,Borowka:2015ura}. Those
differences in the renormalization schemes also affect the comparison
between our results for $M_h^{\text{new}}$ and the results for
$M_h^{\text{old}}+\mathcal{O}(\alpha_t\alpha_s
p^2)+\mathcal{O}(\alpha \alpha_s)$ from Ref.~\cite{Degrassi:2014pfa}
in Tab.~\ref{tab:comparison}.

\begin{table}[t!]
  \centering
  \begin{tabular}{|l|*{4}{S[table-format=3.2]|}}
  \hline scenario & \multicolumn{1}{c|}{$m_h^{\text{max}}$} & \multicolumn{1}{c|}{$m_h^{\text{mod}+}$} & \multicolumn{1}{c|}{$m_h^{\text{mod}-}$} & \multicolumn{1}{c|}{modified light-stop} \\
  \hline 
  $M_h^{\text{old}}$ (GeV)
  & 128.31 & 125.36 & 124.84 & 122.68 \\ 
  $M_h^{\text{old}}$ (GeV)\cite{Degrassi:2014pfa}
  & 128.32 & 125.36 & 124.84 & 122.67 \\ 
  $M_h^{\text{old}}+\mathcal{O}(\alpha_t\alpha_s p^2)$ (GeV)\cite{Borowka:2014wla} 
  & 128.25 & 125.23 & 123.83 & 122.64 \\
  $M_h^{\text{old}}+\mathcal{O}(\alpha_t\alpha_s p^2)$ (GeV)\cite{Degrassi:2014pfa}
  & 127.94 & 124.98 & 123.96 & 122.33 \\ 
  $M_h^{\text{old}}+\mathcal{O}(\alpha_t\alpha_s p^2)+\mathcal{O}(\alpha \alpha_s)$ (GeV)\cite{Degrassi:2014pfa}
  & 128.38 & 125.63 & 124.90 & 122.46 \\
  $M_h^{\text{new}}$ (GeV)
  & 128.53 & 125.75 & 124.85 & 122.61 \\

  \hline 
  \end{tabular}
  \caption{\label{tab:comparison}
  Comparison of the results for the light Higgs-boson mass with
  Ref.~\cite{Degrassi:2014pfa} and Ref.~\cite{Borowka:2014wla} for
  four benchmark scenarios from Refs.~\cite{Carena:2013ytb}
  and \cite{Bagnaschi:2014zla} with $m_A=500$\,GeV and
  $\tan\beta=20$.}

  \vspace*{2ex}
  \capstart
  
  \begin{tabular}{|l|*{3}{S[table-format=3.2]|}}
  \hline & \multicolumn{1}{c|}{this publication} & \multicolumn{2}{c|}{Ref.~\cite{Degrassi:2014pfa}} \\
  & \multicolumn{1}{c|}{$\delta^{(2)}Z_{\mathcal{H}_{i}}^{\text{Ref.~\cite{Borowka:2014wla}}}$}
  & \multicolumn{1}{c|}{$\delta^{(2)}Z_{\mathcal{H}_{i}}^{\text{Ref.~\cite{Borowka:2014wla}}}$}
  & \multicolumn{1}{c|}{$\delta^{(2)}Z_{\mathcal{H}_{i}}^{\text{Ref.~\cite{Degrassi:2014pfa}}}$} \\\hline \hline
  $m_h^{\text{mod}+}$-like
  & \multicolumn{3}{c|}{$M_{\text{SUSY}}=2$\,TeV} \\\hline
  $M_h^{\text{old}}$ (GeV) & 129.38 & 129.38 & 129.38 \\
  $M_h^{\text{new}}$ (GeV) & 129.92 & 129.92 & 129.84 \\\hline
  & \multicolumn{3}{c|}{$M_{\text{SUSY}}=3$\,TeV} \\\hline
  $M_h^{\text{old}}$ (GeV) & 128.63 & 128.63 & 128.63 \\
  $M_h^{\text{new}}$ (GeV) & 129.62 & 129.61 & 129.59 \\\hline \hline
  $m_h^{\text{mod}-}$-like
  & \multicolumn{3}{c|}{$M_{\text{SUSY}}=2$ TeV} \\\hline
  $M_h^{\text{old}}$ (GeV) & 126.92 & 126.92 & 126.92 \\
  $M_h^{\text{new}}$ (GeV) & 127.34 & 127.33 & 127.44 \\\hline
  & \multicolumn{3}{c|}{$M_{\text{SUSY}}=3$\,TeV} \\\hline
  $M_h^{\text{old}}$ (GeV) & 127.02 & 127.02 & 127.02 \\
  $M_h^{\text{new}}$ (GeV) & 127.80 & 127.80 & 127.94 \\\hline
  \end{tabular}
  \caption{\label{tab:msusydependence}
  Values for the lightest Higgs-boson mass in
  the~$m_h^{\text{mod}+}$-like and~$m_h^{\text{mod}-}$-like scenarios
  of Ref.~\cite{Carena:2013ytb}
  using~\mbox{$M_{\text{SUSY}}=2,3$\,TeV} and $m_A=500$\,GeV,
  $\tan\beta=20$.  The results are compared with those provided by the
  authors of Ref.~\cite{Degrassi:2014pfa} for two different
  wave-function renormalization schemes.}
  \vspace{-1.5ex}
\end{table}

The differences in the renormalization schemes and the dependence on
the parameters in the stop sector are further investigated in
Tab.~\ref{tab:msusydependence}. Here the shifts in the light
Higgs-boson mass are shown for SUSY scales of~$2$\,TeV and~$3$\,TeV,
using otherwise the parameters of the $m_h^{\text{mod}+}$ and
$m_h^{\text{mod}-}$ scenarios.  The results for
$M_h^{\text{old}}+\mathcal{O}(\alpha_t\alpha_s
p^2)+\mathcal{O}(\alpha \alpha_s)$ from Ref.~\cite{Degrassi:2014pfa},
where the mass and Yukawa coupling of the bottom quark have been
neglected, are labelled as $M_h^{\text{new}}$ in
Tab.~\ref{tab:msusydependence}.  Two versions of the results from
Ref.~\cite{Degrassi:2014pfa} are shown, one using the renormalization
scheme adopted in Ref.~\cite{Degrassi:2014pfa}
with~\mbox{$\delta^{(2)}Z_{\mathcal{H}_{i}}=\delta^{(2)}Z_{\mathcal{H}_{i}}^{\text{Ref.~\cite{Degrassi:2014pfa}}}$},
and the other using the renormalization scheme of
Ref.~\cite{Borowka:2014wla}, which we have adopted in the present
work,
with~\mbox{$\delta^{(2)}Z_{\mathcal{H}_{i}}=\delta^{(2)}Z_{\mathcal{H}_{i}}^{\text{Ref.~\cite{Borowka:2014wla}}}$}.%
\footnote{We are very grateful to S.~de Vita for providing us with those
results.}  It can be seen in Tab.~\ref{tab:msusydependence} that there
is very good agreement, at the level of about 10\,MeV, between our
results and the results from Ref.~\cite{Degrassi:2014pfa} using the
renormalization scheme of Ref.~\cite{Borowka:2014wla}. The different
choices of renormalization schemes in the result of
Ref.~\cite{Degrassi:2014pfa} amount to mass shifts of up to 150\,MeV
for the displayed examples.  The difference between $M_h^{\text{new}}$
and $M_h^{\text{old}}$ increases with $M_{\text{SUSY}}$ and reaches up
to~$1$\,GeV for the $m_h^{\text{mod}+}$-like scenario at~3\,TeV.

\tocsubsection{Scenario 1: \texorpdfstring{$m_h^{\text{mod}}$}{m\unicodesubscripth\unicodemodifierm\unicodemodifiero\unicodemodifierd}-like}

In the following we further investigate the numerical impact of our
results, including the effect of non-zero phases of the complex
parameters. We start with an $m_h^{\text{mod}}$-like scenario. The
MSSM model parameters in this scenario are chosen as follows
\begin{align}\label{eq:param_scen1}
&& m_{H^\pm} &= 1.5\,\text{TeV}, & M_2 &= 500\,\text{GeV}, & \lvert M_3\rvert & = 2.5\,\text{TeV},\nonumber\\
\span\span m_{\{\tilde{t},\tilde{b}\}_{\text{L}}} = m_{\tilde{Q}_3} &= 2.1\,\text{TeV}, & m_{\{\tilde{t},\tilde{b}\}_{\text{R}}} & = 2\,\text{TeV}, & \lvert X_t\rvert &= 1.3\,m_{\tilde{t}_{\text{R}}}, & \lvert A_b\rvert &= \lvert A_t\rvert,\nonumber\\
\span\span m_{\{ \tilde{q},\tilde{l} \}_{\{ \text{L},\text{R} \}}} & = 2.5 \,\text{TeV}, & A_{\{q,l\}} &= 0, & q &\in {u,d,s,c}, & l &\in {e,\mu,\tau} \text{ .}
\end{align}
Compared to the original $m_h^{\text{mod}}$ scenario we choose larger
bilinear soft-breaking parameters for the sfermions, and also larger
absolute values for~$\mu$ (see below) and~$M_2$.
Thereby~$m_{\tilde{Q}_3}$ is slightly different
from~$m_{\{\tilde{t},\tilde{b}\}_{\text{R}}}$ in order to avoid
numerical instabilities by degeneracies. However, the general feature
of this scenario is kept: it allows for a wide range of~$X_t = A_t^*
- \mu/\tan\beta$ to be in agreement with experimental bounds. With our
choice of parameters, $A_t$ and~$A_b$ are not expected to be affected
by constraints from charge- and color-breaking
minima~\cite{Frere:1983ag,Gunion:1987qv,Casas:1995pd,Hisano:2006mj,Hisano:2007cz,Hisano:2008hn,Camargo-Molina:2013qva,Hollik:2016dcm}. As~$A_\tau$
has negligible impact on the Higgs mass prediction, we set it to zero.

\begin{figure}[b!]
\centering
\includegraphics[width=.49\textwidth]{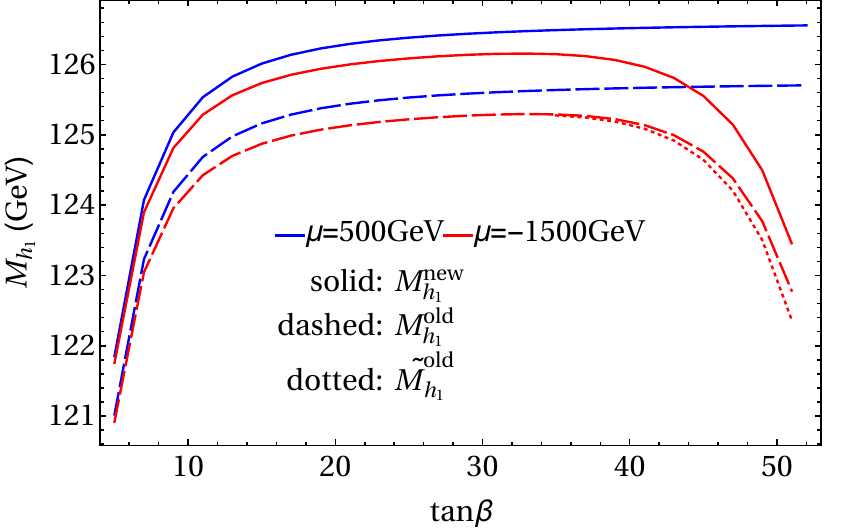}\hfill
\includegraphics[width=.49\textwidth]{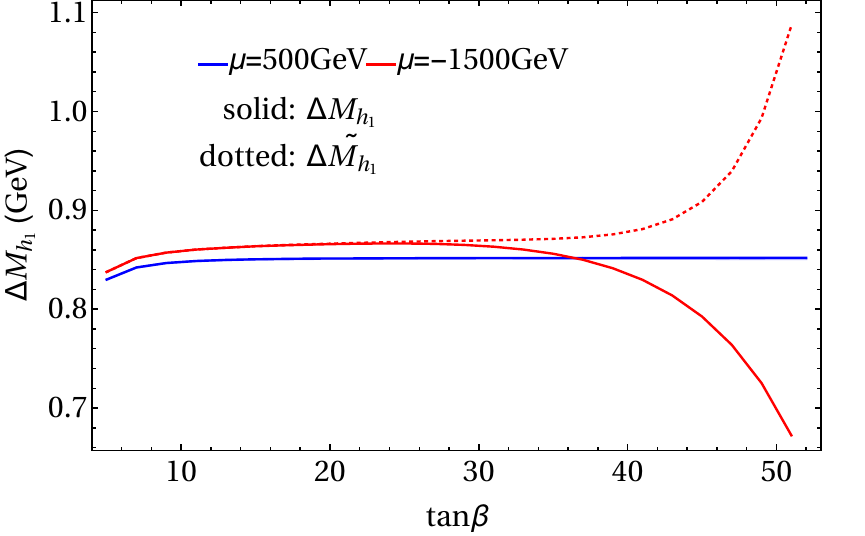}
\caption{\label{fig:scen1:shiftswithtbma0} Prediction for the light
  Higgs-boson mass $M_{h_1}$ (left) and the mass shifts $\Delta
  M_{h_1}$, $\Delta \tilde{M}_{h_1}$ (right, as defined in
  Eq.~(\ref{eq:deltamhdef})) as a function of $\text{tan}\beta$ using
  $m_A$ as input mass for different values of $\mu$. Parameters are as
  described in \eqref{eq:param_scen1}.}

\vspace{7ex}
\capstart

\includegraphics[width=.49\textwidth]{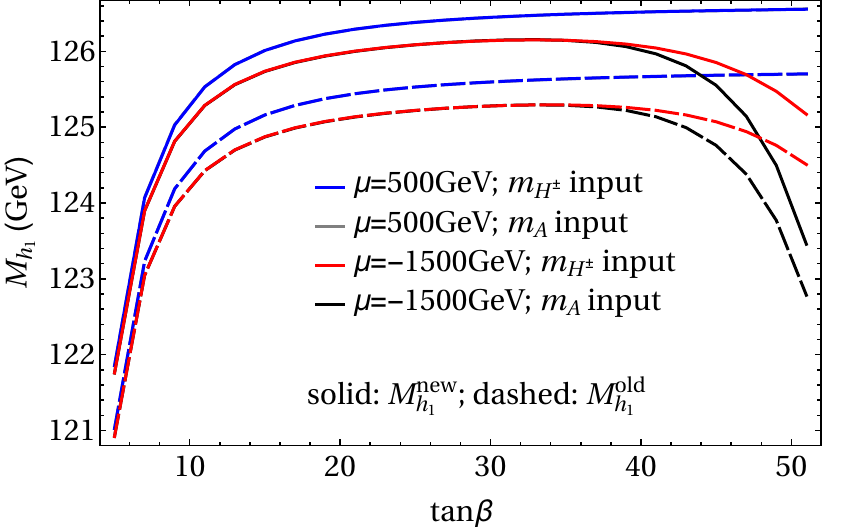}\hfill
\includegraphics[width=.49\textwidth]{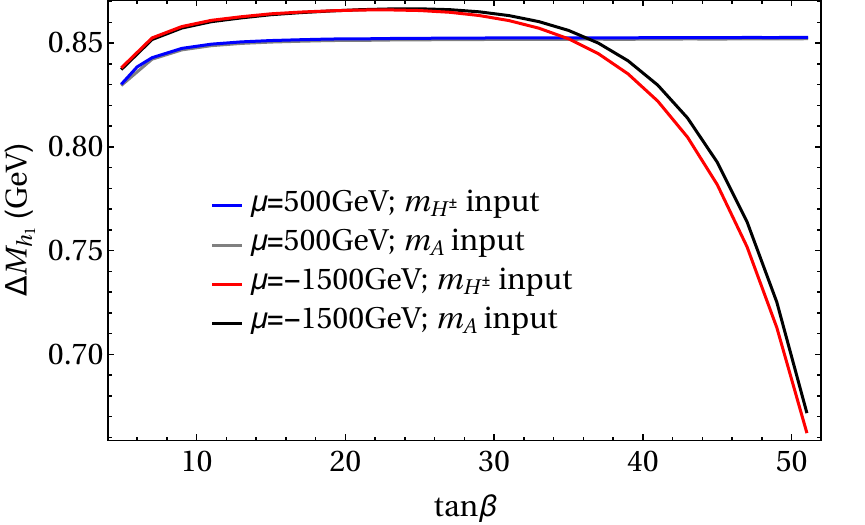}
\caption{\label{fig:scen1:shiftswithtbmhp} Prediction for the light
  Higgs-boson mass ${M}_{h_1}$ (left) and the mass shifts $\Delta
M_{h_1}$, $\Delta \tilde{M}_{h_1}$ (right, as defined in
Eq.~(\ref{eq:deltamhdef})) as a function of $\text{tan}\beta$ using
$m_{H^\pm}$ as input mass for different values of $\mu$.  The black
lines show the results of Fig.~\ref{fig:scen1:shiftswithtbma0} for
$\mu=-1500$\,GeV.  The results of Fig.~\ref{fig:scen1:shiftswithtbma0}
for $\mu=500$\,GeV are indicated by grey lines, which are underneath
the blue lines.  Parameters are as described in
Eq.~\eqref{eq:param_scen1}.}
\end{figure} 

First, the dependence of the lightest Higgs-boson mass~$M_{h_1}$ on
$\tan\beta$ is analyzed for different values of the $\mu$
parameter. Setting all phases of the parameters that can be complex to
zero, our result can be compared to previous ones in the MSSM with
real parameters where the corrections evaluated in the present paper
were not included. In the considered scenario, it is possible to
choose either~$m_A$ or~$m_{H^\pm}=\sqrt{m_A^2 + M_W^2}$ as an input
parameter which is renormalized on-shell accordingly. The chosen input
mass for Fig.~\ref{fig:scen1:shiftswithtbma0} is~$m_A$.  A comparison
of the predicted mass from \texttt{FeynHiggs-2.12.0}, with
($M_{h_1}^{\text{new}}$) and without ($M_{h_1}^{\text{old}}$)
incorporating our new corrections is shown. Solid lines depict the
new, dashed lines the previous results. In order to illustrate the
different relative sizes of our new contributions, we further plot
$\tilde{M}_{h_1}^{\text{old}}$, where the
\texttt{FeynHiggs} result for $M_{h_1}^{\text{old}}$ is
supplemented with
the~$\mathcal{O}{\left(\alpha_b\alpha_s\right)}$ terms known in the
MSSM with real parameters (dotted lines).
The prediction
with~$\mu=500$\,GeV is shown in blue, while the resulting Higgs-boson
mass using~$\mu=-1500$\,GeV is shown in red. The blue dashed and blue
dotted lines are lying on top of each other which means that
the~$\mathcal{O}{\left(\alpha_b\alpha_s\right)}$ corrections are
negligible in this case. 
The red curves show that our new corrections
are significantly larger than the 
pure~$\mathcal{O}{\left(\alpha_b\alpha_s\right)}$ contributions and enter
with different sign. They therefore
overcompensate the slight downward shift induced by the
pure~$\mathcal{O}{\left(\alpha_b\alpha_s\right)}$ contributions. The
differences~$\Delta M_{h_1}$ and~$\Delta \tilde{M}_{h_1}$, as defined
in Eq.~(\ref{eq:deltamhdef}), are plotted on the right-hand side of
Fig.~\ref{fig:scen1:shiftswithtbma0}. For low values
of~$\text{tan}\beta$ the new corrections slightly increase and then
stay constant over a wide range.  Only for values~$\tan \beta>40$ and
large negative~$\mu$ they drop by about~$20\%$. Values for~$\tan
\beta$ above the depicted range and large negative~$\mu$ lead to a
further rapid decrease of~$M_{h_1}$, eventually yielding a tachyonic
Higgs boson. This is due to the large bottom Yukawa coupling with
resummed~$\tan\beta$-enhanced terms which can become non-perturbative
in that region of the parameter space. The rise of the red dotted
curve at large~$\tan\beta$ reflects that this decrease happens for
larger values of~$\tan\beta$ once our new corrections are taken into
account.

In Fig.~\ref{fig:scen1:shiftswithtbmhp} the charged Higgs
mass~$m_{H^\pm}$ is used as an input parameter. The latter implies the
occurrence of terms
of~$\mathcal{O}{\left(\sqrt{\alpha_q}\sqrt{\alpha_o}\alpha_s\right)}$
and corresponds to the renormalization scheme compatible with both the
MSSM with real and complex parameters. On the left-hand side of
Fig.~\ref{fig:scen1:shiftswithtbmhp}, the blue~\mbox{($\mu=500$\,GeV)}
and red~\mbox{($\mu=-1500$\,GeV)} lines show the prediction for the
lightest Higgs mass with~(solid) and without~(dashed) our new
contributions. In addition, the solid and dashed curves of
Fig.~\ref{fig:scen1:shiftswithtbma0} are indicated again as
grey~($\mu=500$\,GeV) and black~($\mu=-1500$\,GeV) lines. In this way,
the influence of the two different renormalization schemes on the
Higgs-mass prediction can be seen. While the blue and grey lines lie
on top of each other over the whole range of~$\tan\beta$, deviations
of up to~$1.5$\,GeV can be observed between the red and black curves
in the region of large $\tan\beta$.  Since the slope of the red curves
for large~$\tan\beta$ is smaller than for the black curves, the
renormalization scheme with~$m_{H^\pm}$ as input parameter is better
suited for this particular region in parameter space. On the
right-hand side of Fig.~\ref{fig:scen1:shiftswithtbmhp} the mass
shifts $\Delta M_{h_1}$ and $\Delta \tilde{M}_{h_1}$ resulting from
our new contributions are depicted. The color coding is the same as
described before. The size of the shifts is almost invariant under the
exchange of~$m_A$ and~$m_{H^\pm}$ as input parameter, since only small
differences between the two renormalization schemes can be noticed.

We note that setting~$\mu=1500$\,GeV and using~$m_{H^\pm}$ as input,
the same qualitative behavior as for the lower positive~$\mu$ value
can be observed, with the new contributions being of the same size as
for~$\mu=-1500$\,GeV in the low and intermediate~$\tan\beta$ region.
Furthermore, the size of the mass shift~$\Delta M_{h_1}$ in
Figs.~\ref{fig:scen1:shiftswithtbma0}
and~\ref{fig:scen1:shiftswithtbmhp} shows a similar tendency with
respect to the chosen sfermion masses as depicted in
Tab.~\ref{tab:msusydependence}, \IE~larger scales increase the size of
the new corrections. However, for stop- and sbottom masses larger
than~$\approx 2$\,TeV logarithmic contributions of higher order also
become important. Then, a resummation of these logarithms should be
taken into account for an accurate Higgs-mass prediction. The gluino
mass can have a sizable impact due to its appearance in the threshold
correction
of~$\mathcal{O}{\left(\alpha_t\alpha_s\right)}$.\footnote{In the
scenario of Eq.~\eqref{eq:param_scen1} the~NNLL-resummation of
logarithms with stop masses as implemented in \texttt{FeynHiggs} gives
rise to an upward shift of the lightest Higgs mass~$M_h^{\text{old}}$
by~$\approx 1.5$\,GeV over the whole range of~$\tan\beta$; for
stop-mass scales at~$3$\,TeV this shift already amounts to~$\approx
2.7$\,GeV. Changing the gluino mass from~$2.5$\,TeV to~$4$\,TeV
corresponds to a downward shift of the SM-like Higgs mass due to the
threshold corrections of~$\mathcal{O}{\left(\alpha_t\alpha_s\right)}$
of~$\approx 0.5$\,GeV over the whole range of~$\tan\beta$.}

\tocsubsection{Scenario 2: large \texorpdfstring{$\tan\beta$}{tan \unicodebeta}}

Scenarios with large values of~$\tan\beta$ are particularly
interesting for investigating effects of the new contributions in the
bottom and sbottom sector.  In that parameter region, terms
proportional to the bottom Yukawa coupling can be as important as
terms from the top sector. In the following, we investigate the
dependence of the new contributions on various parameters at a fixed
large~$\tan\beta$ value.  In order to be consistent with experimental
constraints by ATLAS and CMS we choose a sufficiently large value
of~$m_{H^\pm}$~\cite{Khachatryan:2014wca,ATLAS:2016fpj}. If not stated
otherwise, the MSSM model parameters are
\begin{align}\label{eq:param_scen2}
\tan\beta &= 50, & \mu &= -1.5\,\text{TeV}, & m_{H^\pm} &= 1.5\,\text{TeV}, & M_2 &= 500\,\text{GeV}, & |M_3| & = 2.5\,\text{TeV},\nonumber\\
\span\span m_{\{\tilde{t},\tilde{b}\}_{\text{L}}} = m_{\tilde{Q}_3} &= 2.1\,\text{TeV}, & m_{\{\tilde{t},\tilde{b}\}_{\text{R}}} & = 2\,\text{TeV}, & \lvert X_t\rvert &= 1.3\,m_{\tilde{b}_{\text{R}}}, & \lvert A_b\rvert &= \lvert A_t\rvert,\nonumber\\
\span\span m_{\{ \tilde{q},\tilde{l} \}_{\{ \text{L},\text{R} \}}} & = 2.5 \,\text{TeV}, & A_{\{q,l\}} &= 0, & q &\in {u,d,s,c}, & l &\in {e,\mu,\tau} \text{ .}
\end{align}

\begin{figure}[b!]
\vspace{2ex}
\begin{minipage}[t]{.49\linewidth}
\centering
\includegraphics[width=\textwidth]{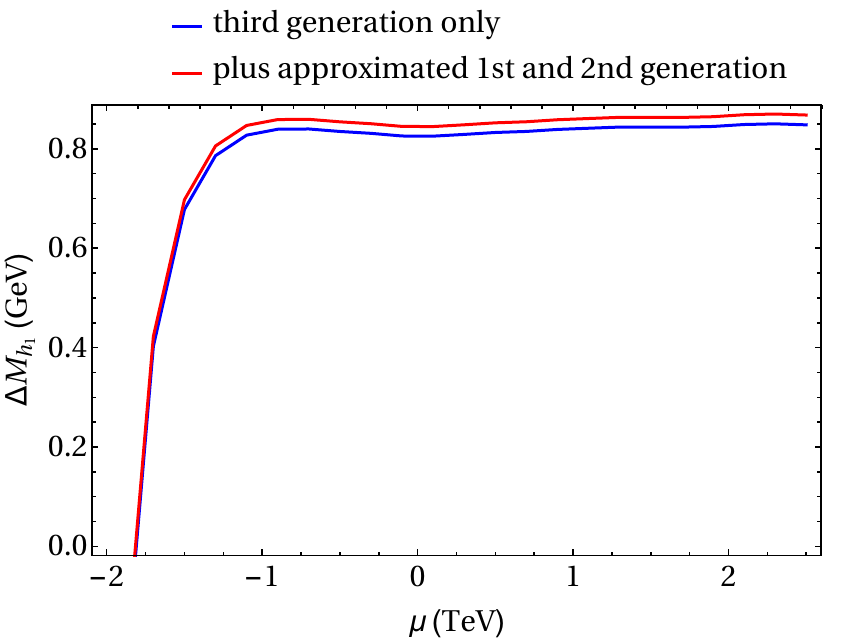}
\caption{\label{fig:scen2:shiftswithmu} Variation of the mass shift
  $\Delta M_{h_1}$ with~$\mu$. The blue curve shows the result including
  contributions only from the 3rd generation. The red line shows the result 
  where also
  contributions of the 1st and 2nd generation are included using the approximation
  $m_q=0,\,q\in\{c,s,u,d\}$. Parameters are as described in
  Eq.~\eqref{eq:param_scen2}.}
\end{minipage}\hfill
\begin{minipage}[t]{.49\linewidth}
\centering
\includegraphics[width=\textwidth]{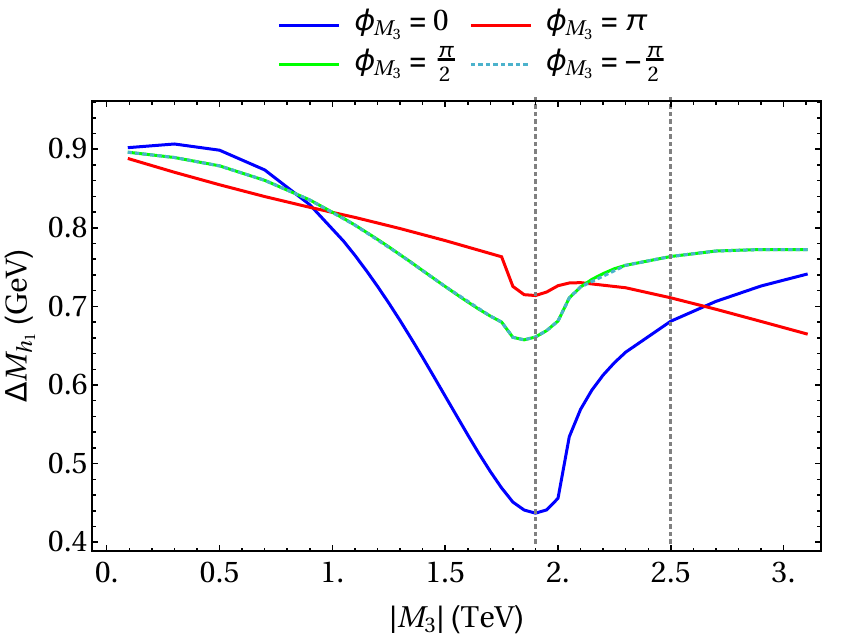}
\caption{\label{fig:scen2:shiftswithM3} Variation of the mass
  shift~$\Delta M_{h_1}$ with the absolute value and phase of the
  gluino mass parameter~\mbox{$M_3 = \lvert M_3\rvert
    \exp{\left(\I\,\phi_{M_3}\right)}$}. The vertical dashed lines are
  at $\lvert M_3\rvert = 1900$\,GeV and $2500$\,GeV. The dependence on
  $\phi_{M_3}$ at those values of $\lvert M_3\rvert$ is illustrated in
  Fig.~\ref{fig:scen2:masswithgluinophase}. Parameters are as
  described in Eq.~\eqref{eq:param_scen2}.}
\end{minipage}
\end{figure}

In Fig.~\ref{fig:scen2:shiftswithmu} the mass shift $\Delta M_{h_1}$
is displayed as a function of $\mu$. Over a wide range the mass shift
is nearly constant at about~$\Delta M_{h_1}\approx 0.85$\,GeV. Only
for large negative values~$\mu\lesssim -1.8$\,TeV, the correction to
the lightest Higgs falls steeply indicating a parameter region where
the perturbative prediction for~$M_{h_1}$ becomes unreliable owing to
the large value of the bottom Yukawa coupling. Thus, $\mu$ should be
kept above that value. The blue line shows the effect of only the
third generation quarks and squarks in our new contributions. The red
line shows the result where these contributions are supplemented with
the corrections of the first and second generation, neglecting the
light quark masses and Yukawa couplings of the first two generations,
$m_q=0,\,q\in\{c,s,u,d\}$. Accordingly, the difference between the two
curves is given by the pure gauge contributions
of~$\mathcal{O}{\left(\alpha\alpha_s\right)}$ from the first and
second generation. They are rather small, amounting to about
$30$\,MeV.

The variation of $\Delta M_{h_1}$ with the gluino-mass parameter $M_3
= |M_3| \exp{\left(\I\,\phi_{M_3}\right)}$ is shown in
Fig.~\ref{fig:scen2:shiftswithM3}. Close to $|M_3|\approx 1.9$\,TeV,
thresholds of the gluino--fermion--sfermion system can be observed,
which are introduced by one-loop integrals entering via the
subloop-renormalization and resummation of the bottom Yukawa
coupling. The effect of varying the absolute value of the gluino-mass
parameter~$\lvert M_3\rvert$ on $\Delta M_{h_1}$ is strongest for
$\phi_{M_3} = 0$ and successively weakened as $\phi_{M_3}$ approaches
$\pi$. The results for $\phi_{M_3} = \pm\frac{\pi}{2}$ almost lie on
top of each other.

\begin{figure}[tp!]
\centering
\includegraphics[width=0.49\textwidth]{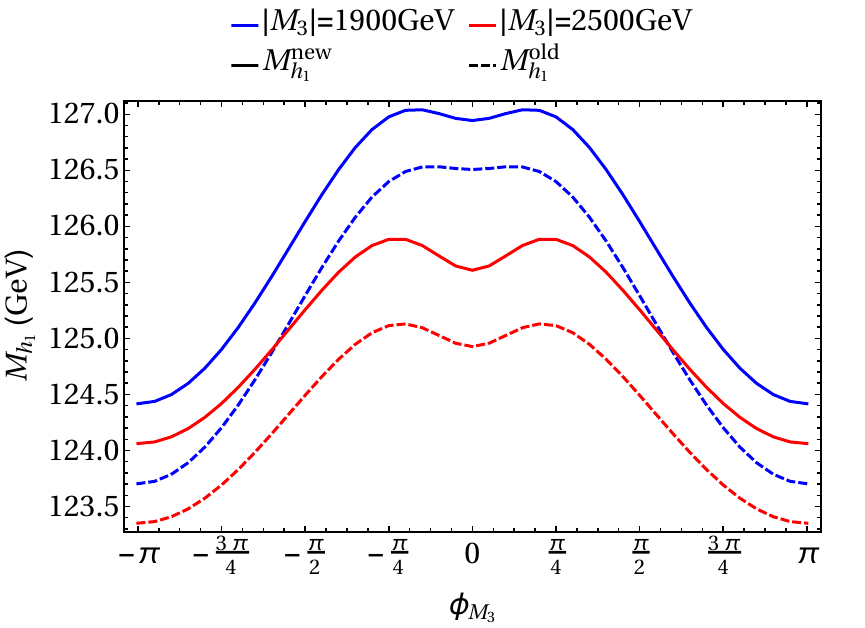}\hfill\includegraphics[width=0.49\textwidth]{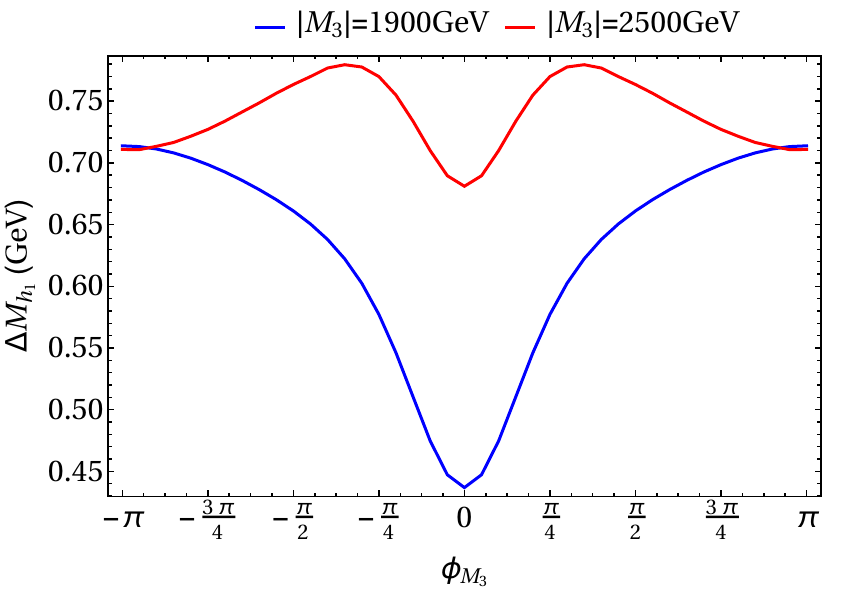}
\caption{\label{fig:scen2:masswithgluinophase} Variation of the light
  Higgs-boson mass $M_{h_1}$ (left) and the 
  mass shift $\Delta M_{h_1}$ (right) with the gluino phase
  $\phi_{M_3}$, while all other phases are 
  set to zero. Parameters are as described in
  Eq.~\eqref{eq:param_scen2}.}

\vspace{5ex}
\capstart

\includegraphics[width=0.49\textwidth]{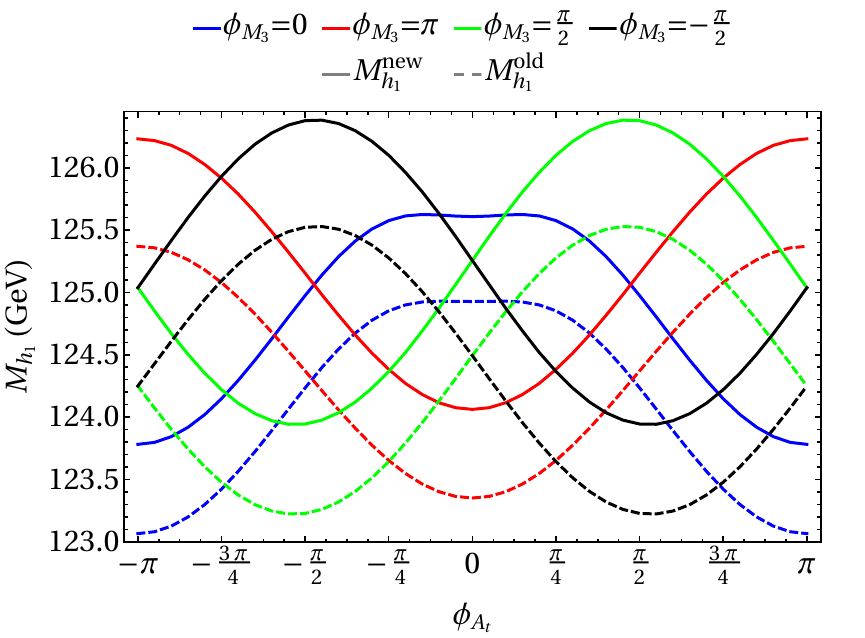}\hfill\includegraphics[width=0.49\textwidth]{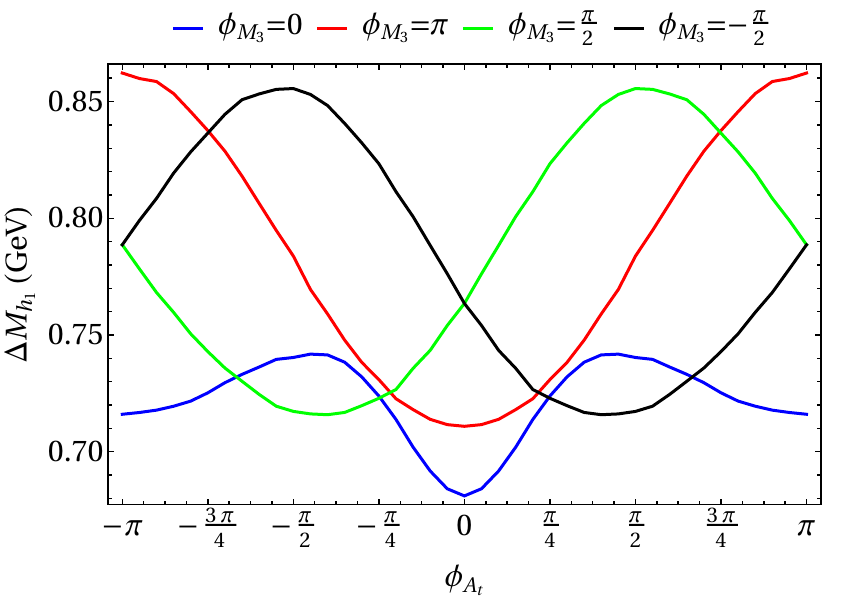}
\caption{
  \label{fig:scen2:masswithAtphase} Variation of the light Higgs-boson
  mass $M_{h_1}$ (left) and the 
  mass shift $\Delta M_{h_1}$ (right) with the phase $\phi_{A_t}$ for different
  $\phi_{M_3}$ and $\phi_{A_b}=0$. Parameters are as described in
  Eq.~\eqref{eq:param_scen2}.}

\vspace{5ex}
\capstart

\includegraphics[width=0.49\textwidth]{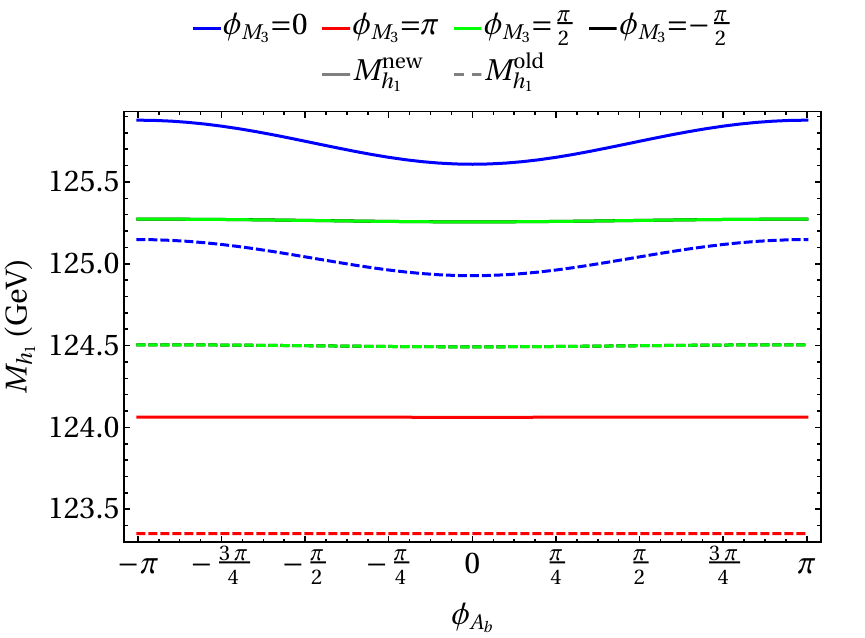}\hfill\includegraphics[width=0.49\textwidth]{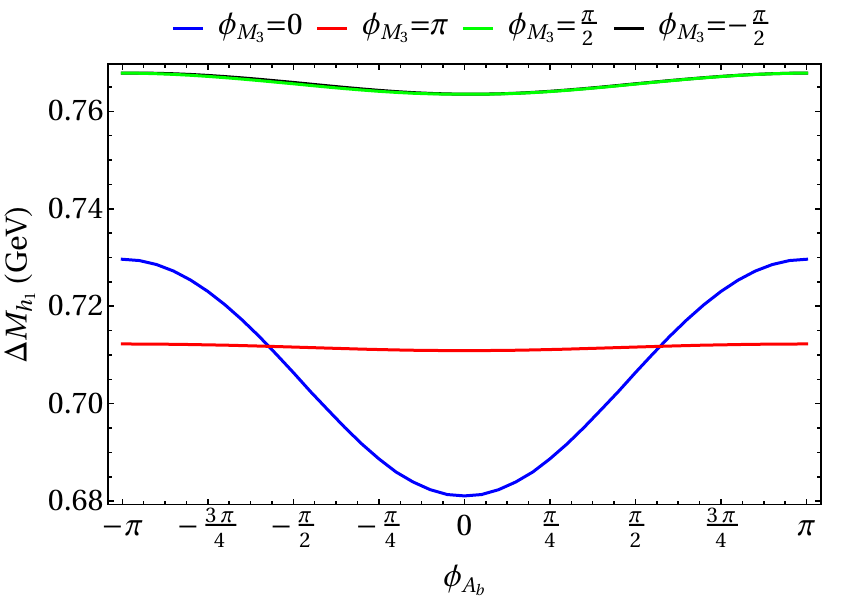}
\caption{\label{fig:scen2:masswithAbphase} Variation of the light
  Higgs-boson mass $M_{h_1}$ (left) and the mass shift $\Delta
  M_{h_1}$ (right) with the phase~$\phi_{A_b}$ for different
  $\phi_{M_3}$ and $\phi_{A_t}=0$. The results for
  $\phi_{M_3}=\pm\frac{\pi}{2}$ lie on top of each other. Parameters
  are as described in Eq.~\eqref{eq:param_scen2}.}
\end{figure}

In
Figs.~\ref{fig:scen2:masswithgluinophase},~\ref{fig:scen2:masswithAtphase}
and~\ref{fig:scen2:masswithAbphase} the dependence on the three
phases~$\phi_{M_3},\,\phi_{A_t}$ and~$\phi_{A_b}$ is displayed,
respectively.  The impact of the new (solid) corrections in comparison
with the ones implemented so far in \texttt{FeynHiggs} (dashed) are
shown for the lightest Higgs-boson mass on the left-hand side of each
figure, while the differences $\Delta M_{h_1}$ are shown on the
right-hand side.  Comparing to the MSSM with real parameters, where
the phases are equal to zero or~$\pi$, sizable differences for the
prediction of the lightest Higgs-boson mass are visible. Concerning
the total variation of $M_{h_1}$ including all now available
corrections, the impact of the phases $\phi_{A_t}$ and $\phi_{M_3}$ is
seen to be rather large with effects that can exceed 2\,GeV, while
varying the phase $\phi_{A_b}$ yields only rather small shifts of
$\approx0.2\,\text{GeV}$.

The prediction for $M_{h_1}$ as function of $\phi_{M_3}$ shown in
Fig.~\ref{fig:scen2:masswithgluinophase} is symmetric with respect to
the sign of $\phi_{M_3}$.  The variation of $\Delta M_{h_1}$ with
$\phi_{M_3}$ is shown on the right-hand side of
Fig.~\ref{fig:scen2:masswithgluinophase}. The pronounced dependence on
the absolute value of $|M_3|$ seen in
Fig.~\ref{fig:scen2:shiftswithM3} can be observed again. The variation
of $\phi_{M_3}$ changes $\Delta M_{h_1}$ by up to $250$\,MeV for an
$|M_3|$ value around the gluino--fermion--sfermion threshold, while
for $|M_3|=2.5$\,GeV $\Delta M_{h_1}$ is shifted only by up to
$70$\,MeV.

The phase dependence of $\Delta M_{h_1}$ on $\phi_{A_t}$ and
$\phi_{A_b}$ is shown on the right-hand side of
Fig.~\ref{fig:scen2:masswithAtphase} and
Fig.~\ref{fig:scen2:masswithAbphase}, respectively. The variation of
$\Delta M_{h_1}$ with $\phi_{A_t}$ and $\phi_{A_b}$ is seen to be
rather small. It reaches up to $150$\,MeV for the phase $\phi_{A_t}$
and up to $50$\,MeV for~$\phi_{A_b}$. It should be noted that the
results for~$\phi_{M_3} = \pm\frac{\pi}{2}$ lie on top of each other
in Fig.~\ref{fig:scen2:masswithAbphase}.  While the variation with
$\phi_{A_b}$ is rather small for any non-zero~$\phi_{M_3}$, the
variation with $\phi_{A_t}$ is minimal for $\phi_{M_3}=0$ and maximal
for $\phi_{M_3}=\pi$.  Using different values of $\phi_{A_b}$ (and
keeping $\phi_{M_3}$ fixed) has only a small effect on the variation
of $\Delta M_{h_1}$ with $\phi_{A_t}$. The corresponding plot is
therefore not shown here.

\tocsubsection{Scenario 3: low \texorpdfstring{$M_H$}{M\_H}}

In the low-$M_H$ scenario the observed SM-like Higgs boson with a mass
of about $125$\,GeV can be identified with the next-to-lightest
neutral $CP$-even Higgs boson of the MSSM, see
Ref.~\cite{Bechtle:2016kui} for a recent update.  We choose the
following MSSM model parameters,
\begin{align}\label{eq:param_scen3}
\tan\beta &= 6.5, & \mu &= 5\,\text{TeV}, & M_2 &= 300\,\text{GeV}, & \lvert M_3\rvert & = 1.5\,\text{TeV},\nonumber\\
m_{\{\tilde{t},\tilde{b}\}_{\{\text{L},\text{R}\}}} &= 750\,\text{GeV}, & m_{\tilde{\tau}_{\{\text{L},\text{R}\}}} &= 500\,\text{GeV}, & m_{\tilde{q}_{\{ \text{L},\text{R} \}}} &= 1.5\,\text{TeV}, & m_{\tilde{l}_{\{\text{L},\text{R}\}}} &= 250\,\text{GeV},\nonumber\\
A_t = A_b = A_{\tau} &= -70\,\text{GeV}, & A_{\{q,l\}} &= 0, & q &\in {u,d,s,c}, & l &\in {e,\mu} \text{ .}
\end{align}
Compared to the original scenario in~\cite{Bechtle:2016kui} we had to
choose a smaller value of~$\mu$ in order to avoid a tachyonic lightest
Higgs boson for a charged Higgs mass~$m_{H^\pm}\approx 160$\,GeV. Our
value for~$\tan\beta$ is chosen such that the scenario is valid
according to Fig.~26 of~\cite{Bechtle:2016kui}.

\begin{figure}[b!]
\centering
\includegraphics[width=0.5\textwidth]{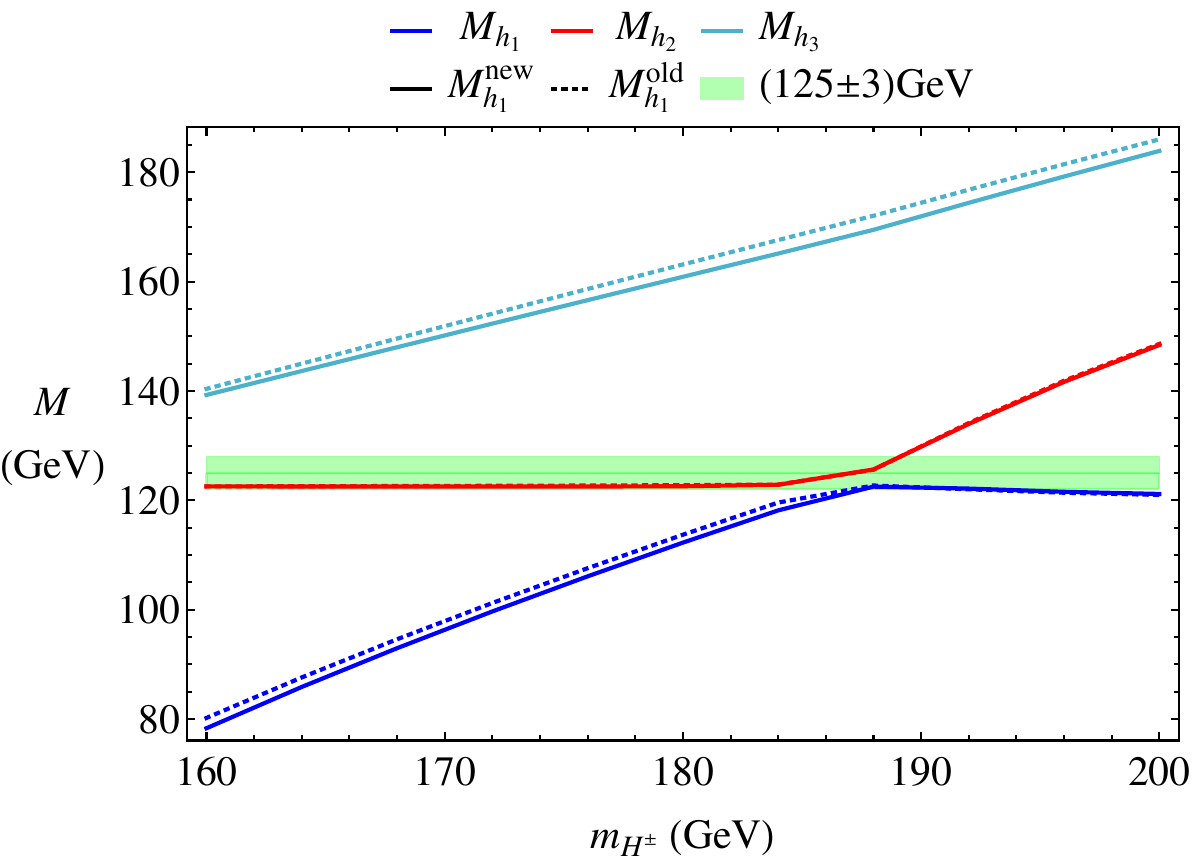}
\caption{\label{fig:scen3:masswithmhpwzerophim3} Variation of the three
  neutral
  Higgs-boson masses $M_{h_i}$ with the charged Higgs boson mass
  $m_{H^\pm}$. The results for $M_{h_i}^{\text{new}}$ are shown as full
  lines and those for
  $M_{h_i}^{\text{old}}$ as dotted lines. Parameters are as described
  in Eq.~\eqref{eq:param_scen3}.}
\end{figure}

\begin{figure}[t!]
\includegraphics[width=\textwidth]{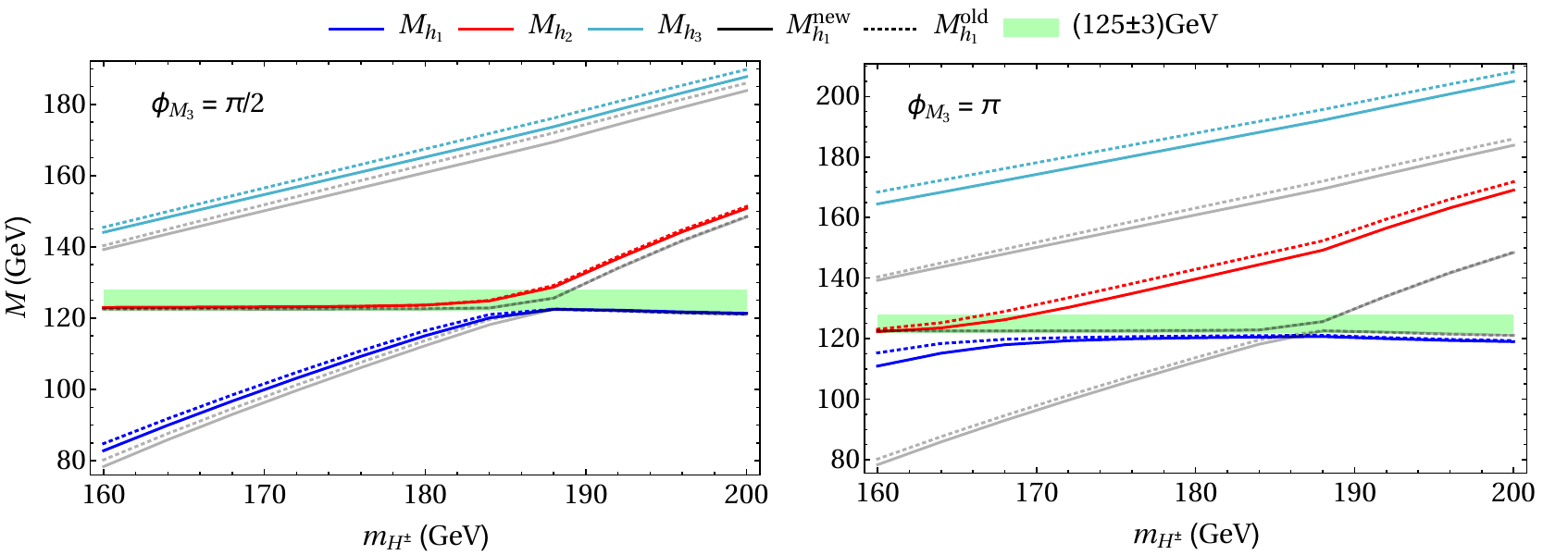}
\caption{\label{fig:scen3:masswithmhp}Variation of the three neutral
  Higgs-boson masses $M_{h_i}$ with the charged Higgs boson mass
  $m_{H^\pm}$ for non-zero phases $\phi_{M_3}$. The results for 
  $M_{h_i}^{\text{new}}$
  are shown as full lines and those for $M_{h_i}^{\text{old}}$ as dotted lines. The
  results of Fig.~\ref{fig:scen3:masswithmhpwzerophim3} with
  $\phi_{M_3} = 0$ are depicted in grey for reference. Parameters are
  as described in Eq.~\eqref{eq:param_scen3}.}
\end{figure} 

In Fig.~\ref{fig:scen3:masswithmhpwzerophim3} the three neutral
Higgs-boson masses are depicted, varying the charged Higgs-boson
mass~$m_{H^\pm}$ which is used as an input parameter. The light green
band illustrates the mass range of~$125\pm 3$\,GeV; it should be
interpreted as a rough indication of the mass range which is
theoretically in agreement with the detected Higgs boson. Up to
$m_{H^\pm}\lesssim 188$\,GeV the heavier Higgs~$h_2$ could be
associated with the discovered Higgs-like particle; however, as can be
seen in the low-$M_H^{\text{alt}+}$ scenario in Fig.~26
of~\cite{Bechtle:2016kui}, our choice of~$\mu$ and~$\tan\beta$ is
already excluded for a charged Higgs mass~$m_{H^\pm}=185$\,GeV. Yet,
scenarios with values of~$m_{H^\pm}$ closer to or below~$m_t$ are
still allowed. In this region the new corrections presented here have
a negligible impact on~$M_{h_2}$, but lead to a downward shift of
about~$1$\,GeV for both $M_{h_1}$ and $M_{h_3}$.

As shown in Fig.~\ref{fig:scen3:masswithmhp}, using a non-zero value
of the gluino phase of $\phi_{M_3} = \pi/2$ or $\phi_{M_3} = \pi$
shifts all three neutral Higgs masses to larger values as compared to
the case $\phi_{M_3}=0$. For better comparison, the results of
Fig.~\ref{fig:scen3:masswithmhpwzerophim3} are underlaid in grey.  The
numerical impact of the new contributions presented here rises with
increasing $\phi_{M_3}$. For $\phi_{M_3}=\pi$ all neutral Higgs masses
can receive large corrections of up to $5$\,GeV.

\tocsection[\label{sec:concl}]{Conclusions}

We have computed the full two-loop QCD corrections to the lightest
Higgs-boson mass in the MSSM with complex parameters. Compared to
previous works, this primarily involves going beyond the gaugeless
limit, and including a finite bottom-quark mass; furthermore the
momentum dependence of loop integrals is taken into account. On the
technical side, this involves the computation of 177 different mass
topologies evaluated at different kinematical configurations,
amounting to a total of 513 two-point two-loop integrals with up to
five mass scales. These integrals have been computed numerically with
the program \texttt{SecDec}.

\medskip

In the first part of our numerical analysis, we have compared our
results with earlier result in the literature taking the appropriate
limit of real parameters and / or vanishing external momentum of our
results. We have found very good agreement with the existing results
in the approriate limit if the same renormalization scheme is
employed. The contributions evaluated in this paper yield a shift in
the lightest Higgs-boson mass at the level of 1\,GeV, where the impact
has been seen to be more pronounced for an increasing mass scale of
the stops.

We have furthermore investigated the dependence of the new corrections
on $\tan \beta$ choosing different values of the $\mu$-parameter as
well as different renormalization schemes. For a large negative $\mu$
the corrections are generally larger and amount to around $~0.9$\,GeV
in $M_{h_1}$. The corrections are largest for $10<\tan \beta<30$,
decrease by $3\%$ for lower values and by about $20\%$ beyond
$\tan \beta=30$.

We find non-vanishing mixed up- and down-type Yukawa corrections in
the charged Higgs-boson self-energy correction entering the mass
predictions for the neutral Higgs bosons as renormalization constant
if the charged Higgs mass $m_{H^\pm}$ instead of the neutral
$\mathcal{CP}$-odd mass $m_A$ is chosen as an input parameter. We have
compared the mass prediction for the lightest Higgs boson in both
schemes and have found good agreement in general. However, using the
charged Higgs mass as an input parameter yields better numerical
stability at large $\tan\beta$ and large negative $\mu$.

The Yukawa contributions scale according to their Yukawa couplings,
leading to much smaller contributions from the first and second
generation quarks and squarks. The pure gauge terms of
$\mathcal{O}{\left(\alpha\alpha_s\right)}$ in the limit of massless
quarks are found to be of similar small size, below $20$\,MeV for one
generation.

Analyzing the dependence on the gluino mass, we have found maximal
shifts of $\approx 900$\,MeV in $M_{h_1}$. The corrections show a
sensitive dependence on the gluino--fermion--sfermion threshold, which
enters via the counterterms of our renormalization scheme, and the
gluino phase.  For the $\mu$-parameter a mass shift of the lightest
Higgs by $\approx 850$\,MeV is found over large regions of parameter
space.

Concerning the impact of the three phases $\phi_{M_3}$, $\phi_{A_t}$
and $\phi_{A_b}$, we find significant effects in our new corrections
from varying the gluino phase and the pase of $A_t$. For $\phi_{M_3}$
the phase dependence becomes particularly pronounced in the threshold
region of the gluino--fermion--sfermion system, as mentioned above.

\medskip

Besides scenarios where the lightest neutral Higgs boson in the
spectrum of the MSSM is the SM-like state that can be identified with
the detected Higgs signal, we have also analyzed the impact of the
newly computed contributions on the Higgs-mass predictions for the
three neutral Higgs bosons within the low-$M_H$ scenario for different
values of the gluino phase $\phi_{M_3}$.  We have found mass
corrections of~$\approx 1$\,GeV for $\phi_{M_3}=0$ and up to $\approx
5$\,GeV for $\phi_{M_3}=\pi$ in this case.

\medskip

Accordingly, we have found that the subleading two-loop contributions
that we have evaluated in this paper yield a shift in the prediction
for the mass of the light SM-like Higgs boson of the MSSM of up to the
level of 1\,GeV.  The size of the correction sensitively depends on
the mass scales of the stops and sbottoms, on the absolute value and
phase of the gluino mass parameter, as well as on the absolute value
and phase of the trilinear coupling in the stop sector (and to a
lesser extent on the trilinear coupling in the sbottom sector).  While
these findings of course have an impact on the remaining theoretical
uncertainties from unknown higher-order corrections, we do not attempt
to provide an improved estimate of the remaining uncertainties
here. Such an improved estimate should be based on a combination of
the fixed-order result considered here with a resummation of
higher-order logarithmic contributions. We leave such an analysis to
future work.

It should be noted in this context that our results for the
corrections of $\mathcal{O}{\left(\alpha\alpha_s\right)}$ beyond the
gaugeless limit cannot be used directly to infer the possible size of
the corresponding contributions
of~$\mathcal{O}{\left(\alpha^2\right)}$ to the Higgs-boson spectrum,
which are unknown up to now. This is due to the fact that the
requirement of a strong coupling in the corrections that we have
evaluated significantly constrains the structure of the contributing
Feynman diagrams, while additional classes of contributions will have
to be taken into account for a full calculation of the corrections of
$\mathcal{O}{\left(\alpha^2\right)}$.

\medskip

The new contributions evaluated in this paper will be made publicly
available in the program
\texttt{FeynHiggs}.

\section*{\tocref{Acknowledgments}}

We are grateful to S.~Di~Vita for providing us with results for a detailed comparison with 
Ref.~\cite{Degrassi:2014pfa}. We thank 
T.~Hahn, S.~Heinemeyer, W.~Hollik and P.~Slavich for
helpful discussions. S.~B.\ gratefully acknowledges financial
support by the ERC Advanced Grant MC@NNLO (340983) and ERC Starting
Grant "MathAm" (39568) during different stages of this project. The
work of S.~P.\
has been supported by the Collaborative Research Center
SFB 676 of the DFG, ``Particles, Strings and the Early Universe'', and by
the~ANR grant ``HiggsAutomator''~(ANR-15-CE31-0002) during different
stages of the project. 
The work of G.~W.\ has been supported in part by the DFG through the SFB 676
``Particles, Strings and the Early Universe'' and by the European Commission
through the ``HiggsTools'' Initial Training Network PITN-GA-2012-316704.

\begingroup
\setstretch{.5}
\bibliographystyle{h-physrev}     
\bibliography{literature}
\endgroup

\end{document}